\documentclass[]{aa}
\usepackage{graphics}
\usepackage{supertabular}
\begin{document}

   \thesaurus{13     
              (13.18.1  
              )}
   \title{Observations of the bright radio sources in the North Celestial Pole
          region at the RATAN-600 radio telescope}


   \author{
          M.G. Mingaliev \inst{1},       
	    V.A. Stolyarov \inst{1,2},
          R.D. Davies \inst{3},
	    S.J. Melhuish \inst{3},
	    N.A. Bursov \inst{1} and
	    G.V. Zhekanis \inst{1}		
	    }

   \offprints{V. Stolyarov}

   \institute{
              Special Astrophysical Observatory of Russian Academy of Sciences,
              Nizhnij Arkhyz, Karachaevo-Cherkessia Republic, 357147, Russia\\
              email: marat@sao.ru
        \and
             University of Cambridge, Institute of Astronomy,
             Madingley Rd.,Cambridge, CB3 OHA, UK\\
             email: vlad@ast.cam.ac.uk
	\and
  	     University of Manchester, Jodrell Bank Observatory,
             Macclesfield, Cheshire, SK11 9DL, UK\\
             email: rdd@jb.man.ac.uk; simon@melhuish.net
}           

   \date{Received November 3, 2000; accepted January 26, 2001}

   \authorrunning{Mingaliev, Stolyarov and Davies}
   \titlerunning{ Observations in the NCP region}

   \maketitle

   \begin{abstract}
A survey of the North Celestial Pole region
using the RATAN-600 radio telescope at five frequencies in the range 
2.3 to 21.7~{\rm GHz} is described. Sources were chosen from the NVSS catalogue. The flux 
densities of 171 sources in the Declination range $+75\degr$ to 
$+88\degr$ are presented; typical flux density errors are 5-10 percent
including calibration errors. About 20 percent of the sources have
flat spectra or a flat component.
   \end{abstract}

%

\section{Introduction}

In the current paper we present the results of observations
of bright radio sources in the North Celestial Pole (NCP) region
within the declination range of $+75\degr\le\delta\le+88\degr$
taken with the RATAN-600 radio telescope of the Russian Academy of Sciences
(Korolkov \& Pariiskii~\cite{korolkov}, Parijskij~\cite{parijskij}).
This NCP survey was initiated as a compliment to the 5~{\rm GHz} interferometric
study of Galactic foreground emission in the NCP made at Jodrell Bank in 1998-1999
(Melhuish et al. \cite{melhuish}). In order to obtain
information about Galactic synchrotron and free-free emission in
the survey area it was necessary to determine the 5~{\rm GHz} flux densities
of the point sources in the area and remove their contribution from
the map.

Up to the present time, there has been no sensitive survey of the NCP region
at frequencies higher than the 1.4~{\rm GHz} NRAO VLA Sky Survey (NVSS),
(Condon et al. \cite{condon}). The 5~{\rm GHz} Greenbank survey (Gregory et al.
\cite{gregory}) only extends northwards as far as $\delta=+75\degr$. There is
limited data available at 5~{\rm GHz} from the early survey at
$+88\degr\le\delta\le+90\degr$ of Pauliny-Toth et al. (\cite{pauliny-toth})
and from the Kuehr et al. (\cite{kuehr}) catalogue of bright sources. Furthermore,
since a sizable fraction (perhaps as many as 20\%) of 5~{\rm GHz} sources may have flat 
spectra and are variable, a contemporary survey covering 5~{\rm GHz} was required
for the NCP project. The upper Declination limit of the present RATAN-600
survey was set at $+88\degr$ since the telescope is used in transit mode
and data cannot be collected close to the NCP in this mode.


\section{Selection criteria for the survey}

The aim of this NCP survey was to obtain information about bright 
point sources which might make a significant contribution to the 5~{\rm GHz} 
degree-scale interferometer survey of foreground Galactic emission
in the NCP (Melhuish et al. \cite{melhuish}). The interferometer has a resolution of
$2\degr$ and a temperature/flux density sensitivity of 60~$\mu {\rm K}$ in antenna
temperature per Jansky. In order to achieve a survey sensitivity approaching 
10~$\mu {\rm K}$ it was decided to measure directly with RATAN-600 all those
sources giving 10~$\mu {\rm K}$ or more with the interferometer, corresponding to a flux density
S~$\ge$ 150~{\rm mJy}. At this flux density there is one source per 
interferometer beam area of $2\degr \times 2\degr$.

The sources chosen for measurement were taken from the 1.4~{\rm GHz} NVSS catalogue,
the catalogue covering the NCP region which is nearest in frequency to 5~{\rm GHz}.
A 150~{\rm mJy} flux density limit at 5~{\rm GHz} corresponds to 350-400~{\rm mJy} at 1.4~{\rm GHz}
assuming an average spectral index of 0.7 ($S \propto \nu^{- \alpha}$) 
for the sources. Accordingly, the
adopted "complete sample" criteria for the sources selected from the NVSS
catalogue were:

\begin{enumerate}
        \item{Flux density $S_{\nu}\ge400~{\rm mJy}$ at the NVSS frequency of 1.4~{\rm GHz}}
        \item{$00^{\rm h}\le\alpha\le24^{\rm h}$}
        \item{$+75\degr\le\delta\le+88\degr$}
\end{enumerate}

In total we have selected for observation 182 objects which satisfy these
criteria.

\section{The observations}

The observations were made in February-March 1999 using the South sector
of the RATAN-600 reflector -- type radio telescope at 2.3, 3.9, 7.7, 11.2 and 21.7~{\rm GHz}
(Parijskij~\cite{parijskij}, Berlin et al.~\cite{berlin}; Berlin \& Friedman~\cite{berlin96}). 
The parameters of the receivers are listed in
Table~\ref{receivers}, where $\nu_{\rm c}$ is the central frequency, $\Delta\nu$ is the
bandwidth, $\Delta$T is the sensitivity of the radiometer over 1 s integration,
T$_{\rm phys}$ is the physical temperature of the radiometer amplifier, T$_{\rm ampl}$ is
the noise temperature of the amplifier and T$_{\rm sys}$ is the noise temperature
of the whole system at the given frequency. All of the radiometers have HEMT first-stage
amplifiers.

\begin{table}
     \caption[]{Parameters of the receivers used in the survey. See text for
meaning of the symbols.}
        \label{receivers}
	{\small
        \begin{tabular}{@{}c|c|c|c|c|c}
        $\nu_{\rm c}$, {\rm GHz} & $\Delta\nu$, {\rm GHz} & $\Delta$T, {\rm mK} & T$_{\rm phys}$, {\rm K}& 
	T$_{\rm ampl}$, {\rm K}& T$_{\rm sys}$, {\rm K}\\
        \hline
        21.7 & 2.5 & 3.5 & 15 & 23 & 77\\
        11.2 & 1.4 & 3.0 & 15 & 18 & 70\\
        7.7  & 1.0 & 3.0 & 15 & 14 & 62\\
        3.9  & 0.6 & 2.5 & 15 & 8  & 37\\
        2.3  & 0.4 & 8.0 & 310& 35 & 95\\
        \hline
        \end{tabular}
	}
\end{table}
%

Information about FWHM can be found in the article by Kovalev et al. (\cite{kovalev}).
For example at 11.2~{\rm GHz} the FWHM is about $17"\times2'$ at the elevations of the NCP
observations.

Usually each source was observed 5-8 times per set. Scans of all of the
sources were corrected for baseline slope when fitted to a Gaussian response
using data reduction software developed by Verkhodanov (\cite{verkhodanov1}).
The accuracy of the antenna temperature of each source was determined as the standard error
of the mean from the $N$ observations of the set.

\section{Calibration and data reduction}

The calibration of our observations is a challenging task.
There are no radio astronomical
calibrators listed in this area of the sky. The only place where
we have some information about source fluxes in a wide
frequency range (0.325 -- 42~{\rm GHz}) in the NCP region is the VLA Calibrator List 
(Perley \& Taylor~\cite{perley}).
However, the fluxes listed there are approximate because most of the
sources from the VLA List are compact and, hence, variable.
To address this problem we selected for our purpose only sources
with steep spectra that are not likely to be variable and, if possible,
with minimal expected VLA amplitude closure errors (about 3\%). 
The fluxes of the calibrators from the VLA Calibrator List are listed at
90, 20, 6, 3.7, 2 and 0.7 cm wavelength bands
(0.325, 1.5, 5, 8.1, 15 and 42.9~{\rm GHz} respectively). In order to get fluxes at the RATAN-600
frequency bands the spectra of the calibrators were interpolated to the desired
frequencies by second order polynomial.

\subsection{The calibration errors, $\sigma_{\rm c}$}

The flux density measurement procedure at the RATAN-600 is described by
Aliakberov et al.~(\cite{aliakberov}).
The response of the antenna to a source with known flux density at a
given frequency $\nu$ is a function of antenna elevation 
(Mingaliev et al.~\cite{mingaliev}), which may be
expressed as:

\begin{equation}
T_{{\rm ant}, \nu} = F_{\nu}(S_{\nu}, e) = S_{\nu}f_{\nu}(e)\\
S_{\nu} =  T_{{\rm ant}, \nu}g_{\nu}(e),
\end{equation}

where
\[
\begin{array}{lp{0.8\linewidth}}
    g_{\nu}(e)=1/f_{\nu}(e) & -  elevation calibration function    \\
    T_{{\rm ant}, \nu}            & -  antenna temperature     \\
    F_{\nu},  f_{\nu}       & -  arbitrary functions     \\
    e = 90 - \phi + \delta  & -  antenna elevation       \\
    \phi = 43\degr.65333    & -  latitude of the telescope site.
\end{array}
\]

   \begin{figure}
   \resizebox{10cm}{!}{\includegraphics{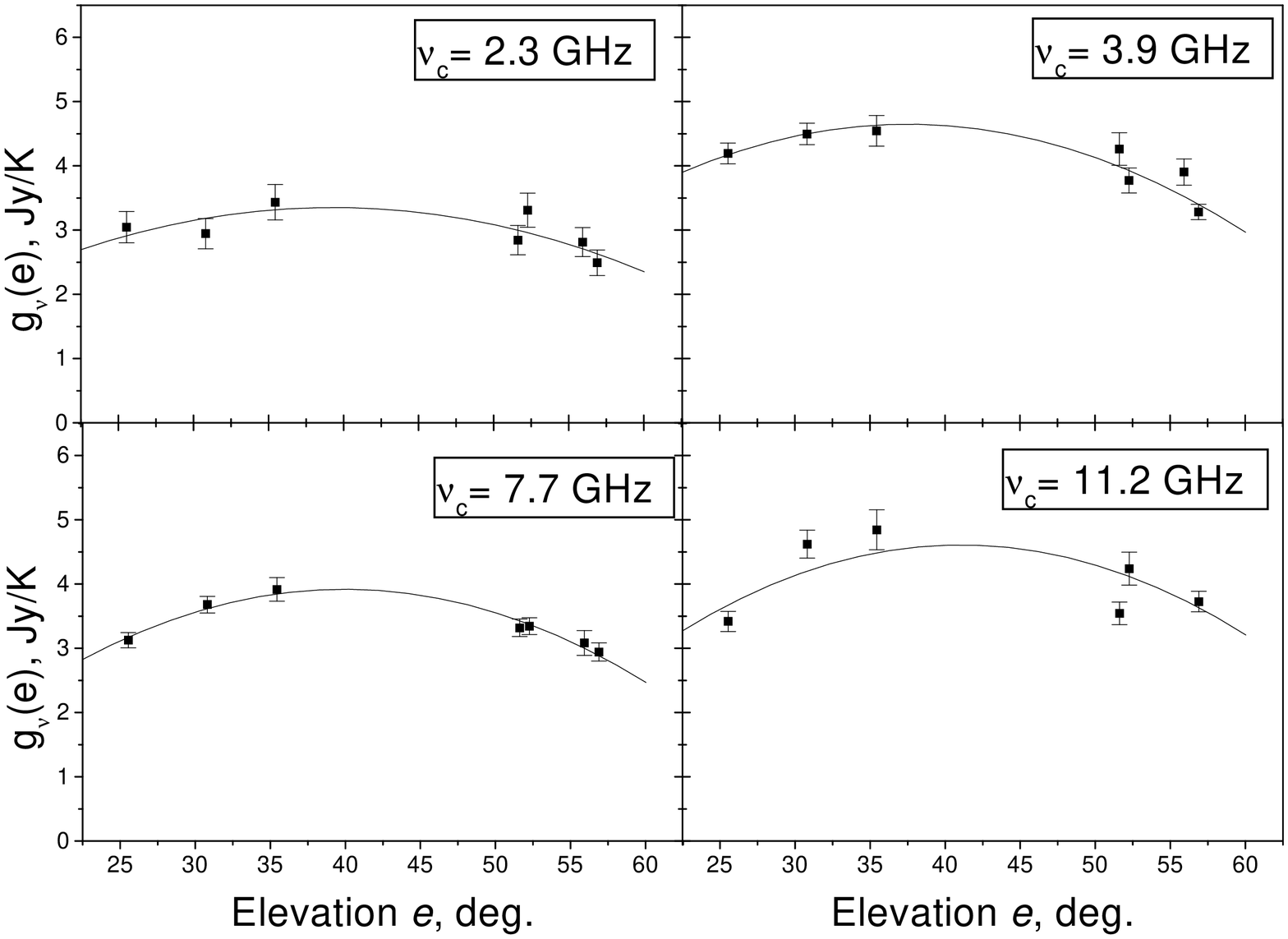}}
   \hfill
   \caption{Calibration curves for 2.3, 3.9, 7.7 and 11.2~{\rm GHz} bands}
   \label{Calibr}
   \end{figure}

In order to get the flux density from $T_{{\rm ant}, \nu}$ we have to multiply
it by the elevation calibration function $g_{\nu}(e)$, which is believed to
be a second order polynomial (Trushkin ~\cite{trushkin}). 
To get an estimate of this function
we observe the calibration sources of known flux density spanning
a wide range of declination, $\delta$. Having the list of values

$g_{\nu}(e_i) = S_{\nu, i}/T_{{\rm ant}, \nu, i}$ \\
for different sources we can approximate the functions $g_{\nu}(e)$
by a second order polynomial 
with the help of minimization of the mean square value of the estimated
error (least square estimator).

The names of calibration sources we used and their adopted flux densities
are listed in the Table~\ref{calibrators}. The assumed flux density
errors are 3\% as given in the VLA Calibrator List.

\begin{table*}
\caption{Adopted calibrator source flux densities. Sources are from the
VSA Calibrator List; errors are assumed to be 3\%}
\label{calibrators}
\begin{tabular}{@{}l|c|c|c|c|c}
Name        &   S($\nu_{\rm c}$=21.7~{\rm GHz}), &   S($\nu_{\rm c}$=11.2~{\rm GHz}),  &  S($\nu_{\rm c}$=7.7~{\rm GHz}),  &   S($\nu_{\rm
c}$=3.9~{\rm GHz}),  &   S($\nu_{\rm c}$=2.3~{\rm GHz}),  \\
of the source & Jy                 & Jy                     & Jy                   &   Jy                 &   Jy                 \\
\hline
J0017+815   &    0.48  &    0.78  &  0.94 &   1.00 &   0.9  \\
J0229+777   &    --    &    --    &  0.41 &   1.05 &   1.8  \\
J0410+769   &    1.14  &    1.77  &  2.23 &   3.35 &   4.49 \\
J0626+820   &    0.31  &    0.56  &  0.74 &   1.01 &   0.95 \\
J1435+760   &    --    &    0.31  &  0.44 &   0.74 &   1.03 \\
J1459+716   &    1.00  &    1.72  &  2.27 &   3.69 &   5.33 \\
J2022+761   &    0.42  &    0.41  &  0.43 &   0.44 &   0.46 \\
\hline
\end{tabular}
\end{table*}

The calibration curves $g_{\nu}(e)$ for 2.3, 3.9, 7.7 and 11.2~{\rm GHz}
are given in Fig.~\ref{Calibr}. A second order polynomial fit was made
to the observational data at each frequency. We found the errors
in the calibration curves were 11, 10.3, 2.4, 5.6 and 7.4 \%
at 21.7, 11.2, 7.7, 3.9 and 2.3~{\rm GHz} respectively. The total calibration
error is the quadratic addition of the 3\% VLA Calibration List error
and the error from the $g_{\nu}(e)$ calibration curve.

\subsection{The errors of $T_{\rm ant}$ measurements, $\sigma_{\rm m}$}

The specifics of the RATAN-600 observations lead to the fact that the errors
of the antenna temperature measurements depend not only on the receiver
noise, but also on the atmospheric fluctuations on the scale of the
main beam, on the 
accuracy of antenna surface setting for the actual
source observation and on the accuracy of the feed cabin positioning (the cabin
with secondary mirror and receivers).

Generally speaking, the part of these errors due to the receiver noise
can be estimated according to the formula

\begin{equation}
\Delta T_{\rm rec} = \Delta T/(\Delta t N k)^{1/2}
\end{equation}  
where $\Delta T$ is the sensitivity of the receiver over 1 s (listed
in Table~\ref{receivers}); 
$\Delta t$ is an integration time, the time that the source takes to 
cross the main beam of the
antenna during the drift scan; $N$ is the number of the drift scans;
$k$ is equal to 1 for single horn receivers (2.3 and 3.9~{\rm GHz}) and
2 for beam-switching receivers (7.7, 11.2 21.7~{\rm GHz}), where we 
can take into account both positive and negative beams. The variable
parameter is $\Delta t$, because it depends on the width
of the main beam which is different for different frequencies, and 
varies with declination. In the case of the NCP declination range
and the frequency range 21.7 to 2.3~{\rm GHz}, $\Delta t$ lies in the range
1 -- 15 s. As an example for 3.9~{\rm GHz}, $\delta = +75\degr$, $\Delta t=4$ s and 
$N$ = 5, $\Delta T_{\rm rec}$ is equal to 0.56 {\rm mK}.

Unfortunately the contribution of atmospheric fluctuations 
increases as $\Delta t$ increases corresponding to larger angular
scales thereby partially reducing the growth of sensitivity expected
from longer integration times.The errors related to the accuracy of antenna 
surface setting and feed cabin positioning are more complicated to account for. 
The feed cabin position errors are most important for high frequency 
observations; as an example it is necessary to position the cabin with an accuracy of
$0.1\lambda$, which is 1.4 mm in the case of 21.7~{\rm GHz}.

However, estimating of the antenna temperature of the source for every
drift scan (e.g by Gaussian fitting) and then calculating the variance
of the $T_{\rm ant}$ for the $N$ observations of the data set can give us 
the measurement error, $\sigma_{\rm m}$, including all of the components listed above.     

\subsection{The total errors, $\sigma_{\rm t}$}

The total fractional error in the flux densities listed in this
survey is the quadratic sum of the total calibration error and the error
in the antenna temperature measurement, namely,

\begin{equation}
(\frac{\sigma_{\rm t}}{S_\nu})^2 = (\frac{\sigma_{\rm c}}{g_\nu(e)})^2 + (\frac{\sigma_{\rm m}}{T_{{\rm ant}, \nu}})^2
\end{equation}

where

\[
\begin{array}{lp{0.8\linewidth}}
    \sigma_{\rm t}              & -  total standard error             \\
    \sigma_{\rm c}              & -  standard error of calibration    \\
    \sigma_{\rm m}              & -  standard error of $T_{{\rm ant},\nu}$ measurement \\
    S_\nu                   & -  flux density            \\
    g_{\nu}(e)=1/f_{\nu}(e) & -  elevation calibration function    \\
    T_{{\rm ant}, \nu}            & -  antenna temperature     
\end{array}
\]

The values of the standard error of $T_{{\rm ant},\nu}$ measurement, $\sigma_{\rm m}$, are 2-3 \% for 11.2, 7,7 and 3.9~{\rm GHz},
3-5\% for 2.3~{\rm GHz} and 7-11\% for 21.7~{\rm GHz}.
For the brighter sources $\sigma_{\rm m}$ is typically half these values,
indicating highly consistent observations.
Thus at the frequencies of 21.7, 11.2 and 3.9~{\rm GHz} the calibration
errors dominate.

\subsection{Comparison with the other catalogues}

   \begin{figure*}
   \resizebox{16cm}{!}{\includegraphics{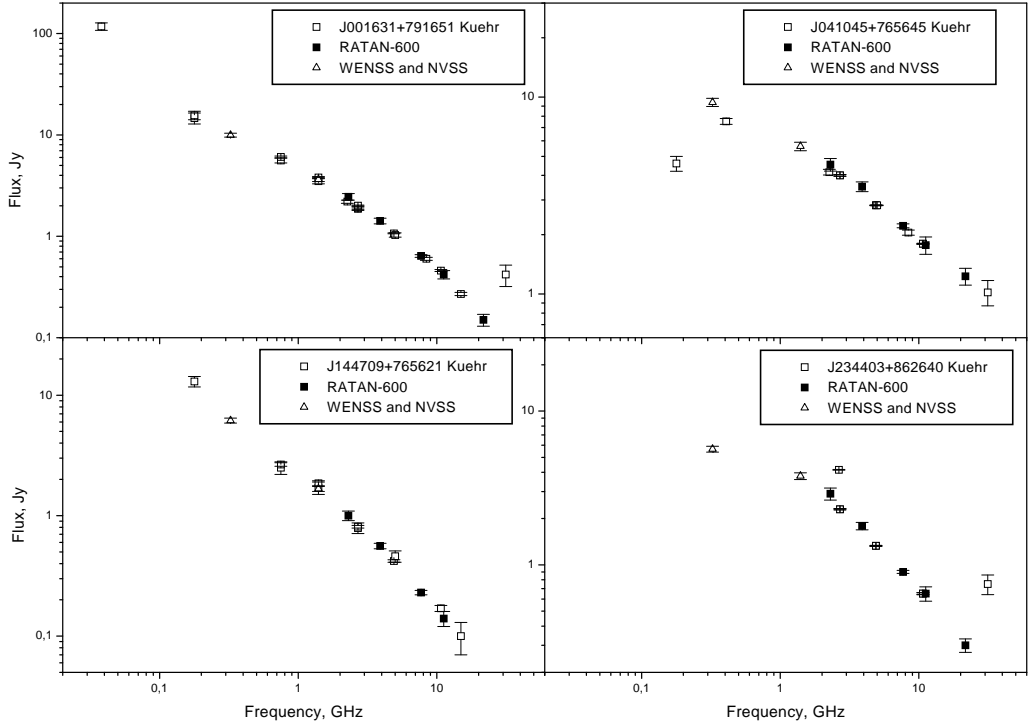}}

   \hfill
   \caption{Four sample source spectra from Kuehr catalogue
            compared with RATAN-600, NVSS and WENSS data}
   \label{Kuehr1}
   \end{figure*}

The RATAN-600 results described in this paper are in good
accordance with the flux densities given by NVSS at 1.4~{\rm GHz},
the Westerbork Northern Sky Survey (WENSS; Rengelink et al.~\cite{rengelink}) 
at 0.325~{\rm GHz} and the earlier data of the Kuehr (\cite{kuehr}) Catalogue.
Four sample spectra are shown in Fig.~\ref{Kuehr1} which compare the
RATAN-600 data with those from the three above catalogues. The sources
illustrated all have steep spectra and as a consequence are not 
likely to be variable. The few discrepancies in the plotted spectra are 
all in the older Kuehr data.

\section{Results}

The spectra for 171 sources in the present RATAN-600 NCP survey are
given in Table~\ref{results}. Data from WENSS and NVSS are included.
Nearly all the sources have complete data at 2.3, 3.9, 7.7 and 11.2~{\rm GHz};
40 sources have 21.7~{\rm GHz} flux densities. Only a few of these sources
have been observed previously over this wide frequency range.
The columns in the Table are:

\[
\begin{array}{lp{0.7\linewidth}}
        Column~1:       &       The source name (NVSS notation), corresponding to epoch J2000 coordinates\\
        Column~2-3:     &       The flux density in {\rm Jy} and standard error at 0.325~{\rm GHz} (WENSS catalogue, 
				Rengelink et al.~\cite{rengelink})\\

        Column~4-5:	&	The flux density in {\rm Jy} and standard error at 1.4~{\rm GHz} (NVSS catalogue,
				Condon et al.~\cite{condon})\\  
	Column~6-15:    &       The flux density in {\rm Jy} and total standard error, $\sigma_{\rm t}$, at 2.3, 3.9, 7.6, 11.2 
	  			and 21.7~{\rm GHz} respectively \\

        Column~16:      &       The spectral index 
                                $\alpha$=-log(S($\nu_{\rm 1}$)/S($\nu_{\rm 2}$))/log($\nu_{\rm 1}$/$\nu_{\rm 2}$),
                                computed between fluxes at 0.325 and 11.2~{\rm GHz} (or the nearest
				available frequencies).
\end{array}
\]

A number of the sources in the NVSS target list were
not fully resolved in the RATAN-600 observations, largely
as a result of the more extended beam in the
declination direction. These closely adjacent sources
are listed as a single entry in Table~\ref{results} and are
designated as RAXXX and DecXXX. The listed flux densities of
these complexes are the sum of the flux densities
of the individual sources. The NVSS sources contributing
to each of the 7 complexes are given in Table~\ref{complex}.

\section{Discussion}

Some preliminary comments are worthwhile on the multifrequency data
for this NCP survey in which 171 individual sources were identified.

\subsection{The contribution to 5~{\rm GHz} interferometry}

The first aim of these observations was to obtain a list of those sources which would 
contribute at a significant level to our 5~{\rm GHz} survey (Melhuish et al.~\cite{melhuish})
of the NCP. The chosen limit to the flux density at 5~{\rm GHz} was 150~{\rm mJy} which corresponds
to a signal amplitude of 10~$\mu {\rm K}$ in the interferometer. The majority of sources
(80 percent) were stronger than this limit and would make a significant contribution
to the CMB foreground and should be removed from the interferometer survey.

The question then arises as to the further contribution from flat spectrum and 
rising spectrum sources not included in our survey which would have a 5~{\rm GHz}
flux density of $\geq$ 150~{\rm mJy}. Remembering that our source selection
criterion was 400~{\rm mJy} at 1.4~{\rm GHz}, a spectral index of 0.7 gives a flux density of
150~{\rm mJy} at 5~{\rm GHz}. A source spectral index of 0.2 will give twice this limit;
only 10 percent of our sources chosen at 1.4~{\rm GHz} have spectral indices flatter than
this value. Accordingly there will be a further contribution from such sources
with flux densities at 1.4~{\rm GHz} of 200 -- 400~{\rm mJy}. Assuming the fraction of flat
spectrum sources stays constant with decreasing frequency, we may expect $\sim$ 5 sources
in this category. Yet another contribution will come from Gigahertz Peaked Spectrum (GPS)
sources; likewise there will be $\sim$ 5 extra sources with a flux density above
150~{\rm mJy} at 5~{\rm GHz}.

\subsection{Statistics of sources spectra}

\begin{figure}
   \resizebox{10cm}{!}{\includegraphics{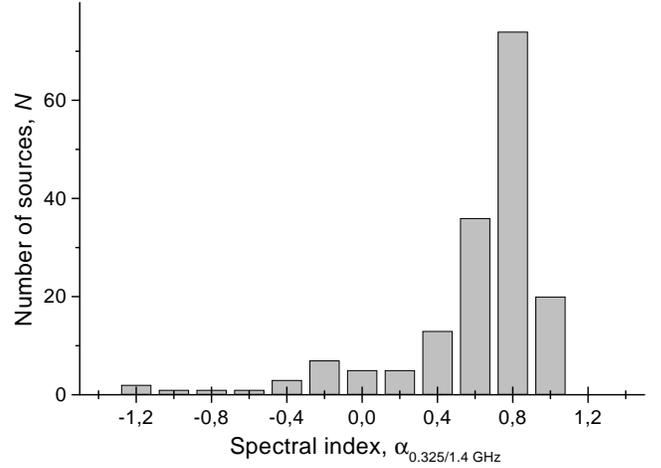}}
   \hfill
   \caption{Spectral index distribution between 0.325 and 1.4~{\rm GHz}}
   \label{s0325-14}
   \end{figure}

\begin{figure}
   \resizebox{10cm}{!}{\includegraphics{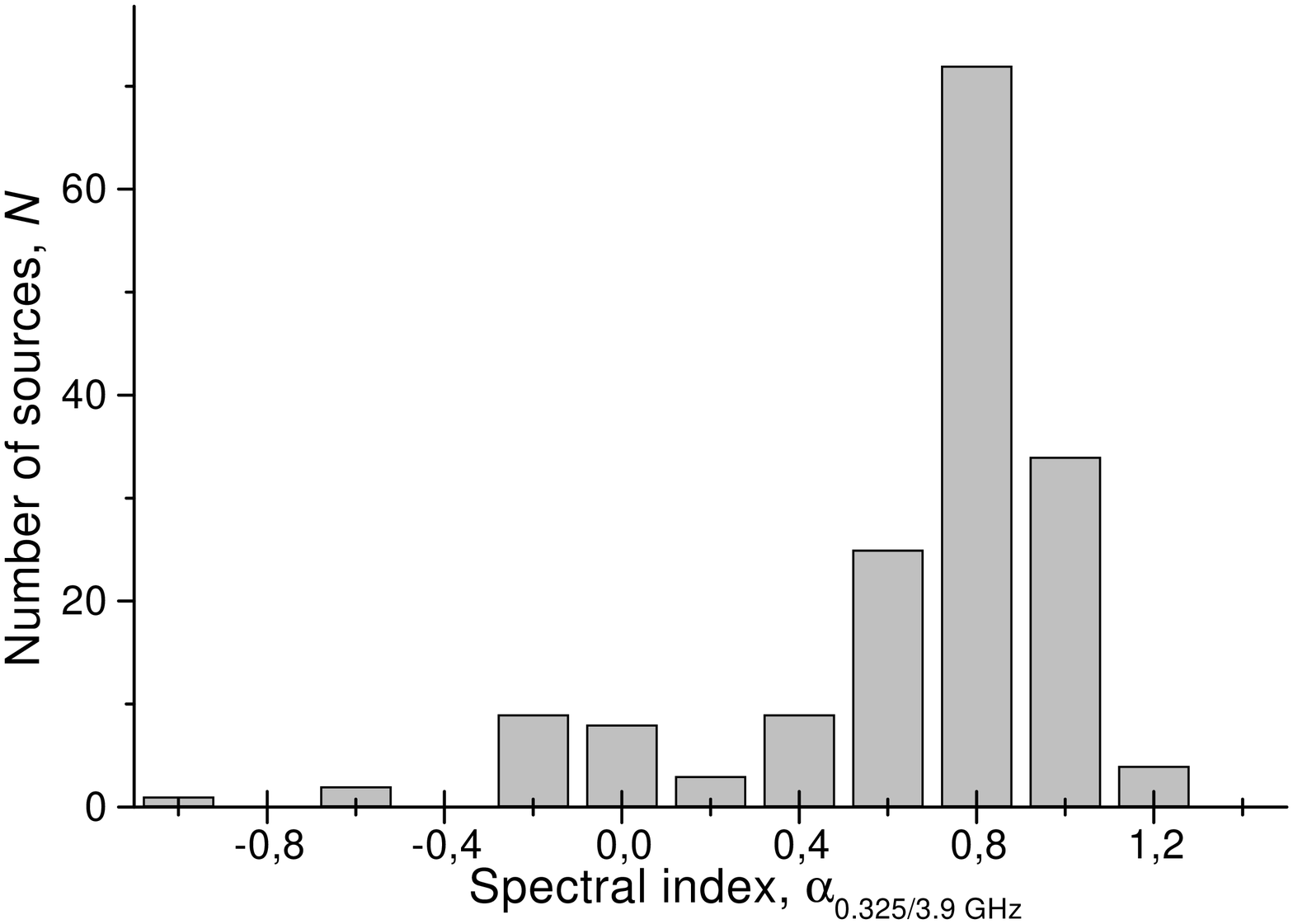}}
   \hfill
   \caption{Spectral index distribution between 0.325 and 3.9~{\rm GHz}}
   \label{s0325-39}
   \end{figure}

\begin{figure}
   \resizebox{10cm}{!}{\includegraphics{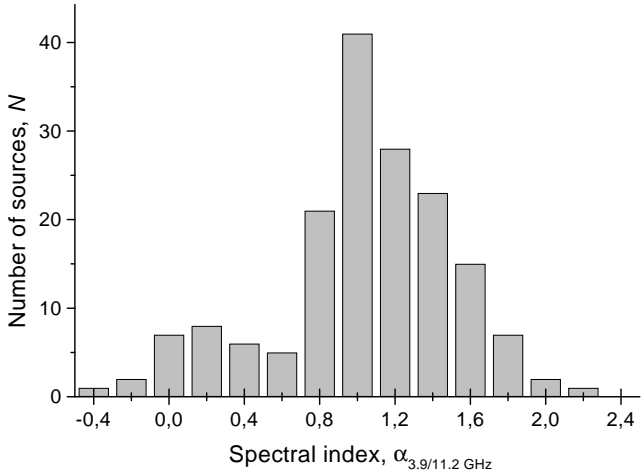}}
   \hfill
   \caption{Spectral index distribution between 3.9 and 11.2~{\rm GHz}}
   \label{s39-112}
   \end{figure}

\begin{figure}
   \resizebox{10cm}{!}{\includegraphics{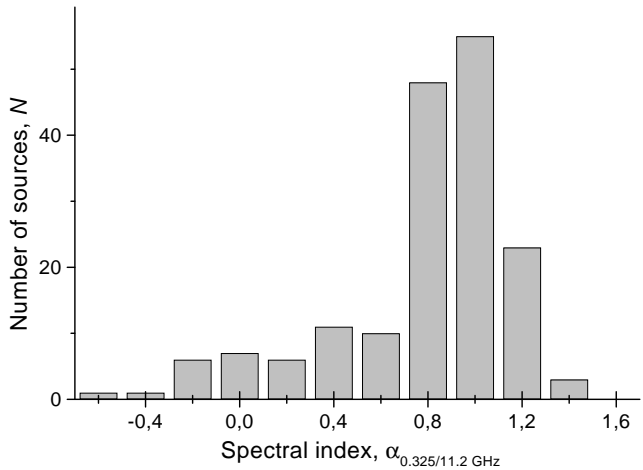}}
   \hfill
   \caption{Spectral index distribution between 0.325 and 11.2~{\rm GHz}}
   \label{s0325-112}
   \end{figure}

Although this is a modest sample of GHz spectra, it provides an indication
of the spectral properties of the brightest radio sources in the NCP region
($+75\degr\le\delta\le+88\degr$). We would expect them to follow the trends
in the general field. One particular advantage of the present catalogue is that
all the sources were observed simultaneously at all frequencies to provide
an instantaneous spectrum unaffected by source variability. The histograms of
spectral index values estimated over the frequency ranges 
0.325/1.4~{\rm GHz}, 0.325/3.9~{\rm GHz}, 3.9/11.2~{\rm GHz} and 0.325/11.2~{\rm GHz} are presented
in Figs.~\ref{s0325-14}, \ref{s0325-39}, \ref{s39-112} and \ref{s0325-112} 
respectively.

The majority of the sources have spectral indices in the range 0.6 to 1.5
at GHz frequencies. The canonical steepening of synchrotron spectra at higher
frequencies is evident in the data. The median spectral index in the
range 0.325/1.4~{\rm GHz} is 0.78. This value rises to 0.82 for 1.4/2.3~{\rm GHz},
to 0.95 for 2.3/3.9~{\rm GHz} and to 1.15 for 3.9/11.2~{\rm GHz}. At the higher part
of this frequency range there is an increasing spread in the range of spectral
indices which indicates that the turn-over of the spectrum occurs at a range
of GHz frequencies. This broadening of the histogram is readily seen in 
Fig.~\ref{s39-112} where a significant number ($\sim$ 30 percent) have a spectral
index greater than 1.2 in the frequency range 3.9/11.2~{\rm GHz}; a small fraction
($\sim$ 10 percent) of these higher spectral indices are a result of the
significant total error, $\sigma_{\rm t}$, on weaker sources at 11.2~{\rm GHz}.

The fraction of flatter spectrum sources in our GHz NCP survey, as illustrated in
Figs.~\ref{s0325-39}, ~\ref{s39-112} and \ref{s0325-112}, is 20 -- 25 percent. This family of 
flattish spectrum sources is of particular concern as a foreground in the
measurement of fluctuations in the cosmic microwave background.

\subsection{Individual sources with compact components}

The spectral signature of compact radio sources is a flat component
arising from synchrotron self-absorption.
Such a component may be seen as a flat spectrum over
a wide frequency range, a flat spectrum component at a high frequency emerging
from a steep spectrum low-frequency source or a GHz Peaked Spectrum source.
Such spectra are found in some 20 percent of the 171 sources of the
present survey. 14 sources show a relatively flat spectrum over the whole
frequency range measured; of these, 4 show weak GPS behavior (see below)
and 3 show weakly rising spectra up to the highest frequency
observed and may also be GPS sources. 4 sources show
evidence for a flat spectrum component at the higher frequencies
of the survey.

There is a potential 10 percent of this survey which
are GPS sources. These are believed to be compact objects with a peak in
their spectra at GHz frequencies in the redshift frame of emission.
They are characterized by a difference of spectral index ("curvature")
on either side of the peak of more than 0.6 
(de Vries et al.~\cite{devries}). The observed peak frequency
may be as low as 0.5~{\rm GHz} (Marecki et al.~\cite{marecki}). There are 9 sources
in our list which satisfy these spectral characteristics. It is possible
that several may be giant radio galaxies with low frequency absorption 
(for example 085834+750121); mapping will be required to establish
their compactness. Four sources (132351+794251; 172359+765312; 180045+782804;
200531+775243) have a curvature of 0.4 to 0.6, just below the canonical
limit of de Vries et al.; they are compact as indicated by their spectra and are 
potential GPS sources. Three sources (104423+805439; 135755+764320; 235622+815252)
have spectra which are weakly rising with $\alpha$ = 0.2 to 0.3 at the highest 
frequencies of observation. These are also likely GPS sources with peak 
frequencies $\ge$ 10 to 20~{\rm GHz}.

\begin{acknowledgements}

This research is partially supported by the Russian Foundation for
Basis Research Project No. 98-02-16428 and Russian Federal Program
"Astronomy" Project No. 1.2.5.1. VS acknowledges the receipt of
NATO/Royal Society Postdoctoral Fellowship.

The authors used extensively the database CATS (http://cats.sao.ru, Verkhodanov
et al., \cite{verkhodanov}) of the Special Astrophysical Observatory (Russia)
in the search for counterparts in the radio catalogues.

\end{acknowledgements}

\begin{center}
\begin{table*}
\caption{Complex sources}
\label{complex}
\small{
\begin{tabular}{@{}l|l|l|l}
\hline 
Name in NVSS	&	Name in the Table~\ref{results} & Name in NVSS	&	Name in the Table~\ref{results} \\
\hline 
J022235+861727  &        J0222XX+861XXX & J184142+794752  &     J184XXX+794XXX  \\
J022248+861851  &			& J184151+794727  &                     \\
J022249+862027  &			& J184214+794613  &			\\	
 		&			& J184226+794517  &			\\
J074246+802741  &        J074XXX+802XXX &		&			\\
J074305+802544  &			& J204257+750428  &     J20425X+750XXX  \\
                &                        &J204259+750306  &			\\
J101330+855411  &        J101XXX+855XXX  &		  &                     \\	
J101412+855349  &                        & J211814+751203 &	J2118XX+751XXX  \\
                &			 & J211817+751112 & 			\\
		&			 & 		  &			\\
		&			&  J235521+795552 &     J2355XX+795XXX  \\
		&			&  J235525+795442 &			\\ 
\hline 
\end{tabular}}
\end{table*}
\end{center}

\begin{center}
\topcaption{Fluxes of the sources in the NCP region}
\label{results}
{\tiny{
\tablehead{\hline
1 & 2 & 3 & 4 & 5 & 6 & 7 & 8 & 9 & 10 & 11 & 12 & 13 & 14 & 15 & 16\\
\hline}
\tabletail{\hline}
\par
\begin{onecolumn}
\begin{supertabular}{c|rr|rr|rr|rr|rr|rr|rr|r}
  Name, &    $S_{\rm 0.325}$ &  $\sigma_{\rm t} $    &    $S_{\rm 1.4}$ &     $\sigma_{\rm t}$ &     $S_{\rm 2.3}$ &     $\sigma_{\rm t}$ &     $S_{\rm 3.9}$ &  
  $\sigma_{\ rmt}$  &     $S_{\rm 7.7}$ &     $\sigma_{\rm t}$ &     $S_{\rm 11.2}$ &     $\sigma_{\rm t}$ &    $S_{\rm 21.7}$ &   $\sigma_{\rm t}$ &     Sp. index\\
  NVSS  &        Jy,     &  WENSS          &        Jy,  &   NVSS          &         Jy   &                &          Jy     &               &         Jy      &                &         Jy &          &     Jy       &       &  $\alpha_{0.325/11.2}$            \\
           &            &            &            &            &            &            &            &            &            &            &            &            &            &            &            \\
\hline
J000943+772440 &      2.059 &      0.082 &      0.628 &      0.021 &      0.447 &       0.04 &      0.267 &      0.017 &      0.161 &      0.007 &       0.14 &      0.016 &          -- &          -- &       0.76 \\
J001236+854313 &      1.605 &      0.064 &      0.691 &      0.023 &      0.539 &      0.048 &      0.246 &      0.018 &      0.148 &      0.005 &      0.099 &      0.011 &          -- &          -- &       0.79 \\
J001311+774846 &      4.517 &      0.181 &      2.101 &      0.071 &      1.578 &      0.141 &      0.981 &      0.056 &      0.467 &      0.012 &      0.314 &      0.034 &      0.202 &       0.03 &       0.75 \\
J001631+791651 &      9.959 &      0.398 &      3.651 &      0.123 &       2.46 &      0.183 &      1.422 &      0.087 &      0.643 &      0.017 &      0.422 &      0.044 &      0.145 &      0.022 &       0.89 \\
J001708+813508 &      0.688 &      0.028 &      0.693 &      0.023 &      0.853 &      0.067 &      1.022 &      0.058 &      0.919 &      0.023 &      0.788 &      0.082 &      0.441 &      0.065 &      -0.04 \\
J001816+782743 &       2.16 &      0.086 &      0.707 &      0.024 &      0.428 &      0.038 &      0.252 &      0.015 &      0.109 &      0.005 &      0.054 &      0.006 &          -- &          -- &       1.04 \\
J003812+844727 &      0.972 &      0.039 &      0.406 &      0.014 &      0.332 &       0.03 &      0.223 &      0.017 &      0.108 &      0.005 &      0.084 &       0.01 &          -- &          -- &       0.69 \\
J004617+751752 &      1.371 &      0.055 &      0.444 &      0.015 &      0.273 &      0.025 &      0.163 &      0.012 &      0.074 &      0.004 &       0.04 &      0.005 &          -- &          -- &          1 \\
J011045+873822 &      2.215 &      0.089 &      0.686 &      0.023 &      0.467 &      0.042 &      0.206 &      0.015 &      0.118 &      0.005 &      0.081 &      0.009 &          -- &          -- &       0.93 \\
J013156+844612 &      2.072 &      0.083 &      0.777 &      0.026 &      0.603 &      0.054 &      0.361 &      0.027 &      0.186 &      0.005 &      0.136 &      0.016 &          -- &          -- &       0.77 \\
J015207+755035 &      2.823 &      0.113 &      0.839 &      0.028 &      0.526 &      0.047 &      0.285 &      0.016 &      0.095 &      0.005 &      0.054 &      0.006 &          -- &          -- &       1.12 \\
J020537+752207 &       0.86 &      0.035 &      1.151 &      0.039 &      0.873 &      0.078 &      0.611 &      0.036 &      0.309 &      0.009 &      0.202 &      0.021 &      0.097 &      0.015 &       0.41 \\
J020723+795602 &       3.89 &      0.156 &      1.363 &      0.046 &      0.932 &      0.072 &      0.582 &      0.033 &      0.252 &      0.008 &       0.18 &      0.021 &          -- &          -- &       0.87 \\
J0222XX+861XXX &     20.538 &      0.822 &      6.478 &      0.217 &      4.006 &      0.422 &      3.006 &      0.261 &      1.137 &      0.133 &      0.813 &      0.095 &          -- &          -- &       0.91 \\
J022454+765554 &      5.949 &      0.238 &      1.927 &      0.065 &      1.271 &      0.095 &      0.744 &      0.042 &      0.334 &      0.009 &       0.21 &      0.023 &      0.085 &      0.013 &       0.94 \\
J022914+774316 &      8.067 &      0.323 &      2.683 &       0.09 &      1.663 &      0.124 &       0.95 &      0.054 &      0.401 &      0.011 &      0.259 &      0.028 &      0.111 &      0.016 &       0.97 \\
J023010+814129 &      2.285 &      0.091 &      1.048 &      0.035 &      0.783 &       0.07 &      0.544 &      0.031 &      0.284 &      0.008 &      0.213 &      0.023 &      0.212 &      0.031 &       0.67 \\
J025100+791359 &      2.475 &      0.099 &      0.713 &      0.024 &      0.475 &      0.043 &      0.226 &      0.013 &      0.089 &      0.003 &      0.036 &      0.004 &          -- &          -- &        1.2 \\
J025417+791147 &      2.186 &      0.088 &      0.643 &      0.022 &      0.463 &      0.041 &      0.212 &      0.013 &      0.105 &      0.006 &      0.067 &      0.008 &          -- &          -- &       0.98 \\
J030011+820238 &      2.415 &      0.097 &      1.379 &      0.046 &      1.078 &      0.102 &      0.586 &      0.034 &      0.265 &      0.008 &      0.177 &      0.022 &          -- &          -- &       0.74 \\
J030454+772731 &      0.164 &      0.008 &      0.977 &      0.033 &      0.926 &       0.07 &      0.634 &      0.036 &      0.344 &      0.009 &      0.262 &      0.028 &      0.157 &      0.024 &      -0.13 \\
J035150+800437 &      1.534 &      0.061 &      0.546 &      0.018 &      0.368 &      0.033 &      0.211 &      0.012 &      0.078 &      0.004 &      0.041 &      0.005 &          -- &          -- &       1.02 \\
J035446+800929 &      1.243 &       0.05 &      0.644 &      0.022 &      0.499 &      0.045 &      0.371 &      0.023 &      0.279 &      0.008 &      0.259 &      0.028 &      0.211 &      0.031 &       0.44 \\
J035629+763742 &      1.969 &      0.079 &      0.629 &      0.021 &      0.397 &      0.036 &      0.252 &      0.016 &      0.127 &      0.007 &      0.098 &      0.011 &          -- &          -- &       0.85 \\
J035817+783719 &      1.334 &      0.053 &      0.508 &      0.017 &      0.367 &      0.033 &      0.226 &      0.016 &      0.102 &      0.003 &      0.065 &      0.007 &          -- &          -- &       0.85 \\
J040652+763354 &      1.592 &      0.064 &      0.527 &      0.018 &      0.384 &      0.034 &      0.195 &      0.012 &      0.095 &      0.004 &      0.068 &      0.008 &          -- &          -- &       0.89 \\
J041045+765645 &      9.406 &      0.376 &       5.62 &      0.189 &      4.538 &      0.336 &      3.512 &      0.197 &      2.215 &      0.054 &      1.769 &      0.182 &      1.279 &      0.142 &       0.47 \\
J041426+761243 &      1.359 &      0.054 &      0.405 &      0.014 &      0.259 &      0.023 &      0.147 &      0.011 &      0.061 &      0.003 &       0.04 &      0.005 &          -- &          -- &          1 \\
J041531+842457 &      1.965 &      0.079 &      0.578 &      0.019 &      0.453 &      0.041 &      0.235 &      0.015 &      0.128 &      0.004 &       0.09 &       0.01 &          -- &          -- &       0.87 \\
J041946+755915 &      1.491 &       0.06 &      0.427 &      0.014 &      0.273 &      0.025 &      0.166 &       0.01 &      0.076 &      0.004 &      0.045 &      0.005 &          -- &          -- &       0.99 \\
J042205+762708 &       3.75 &       0.15 &      1.043 &      0.035 &      0.598 &      0.053 &      0.309 &      0.018 &        0.1 &      0.006 &       0.05 &      0.006 &          -- &          -- &       1.22 \\
J042408+765341 &      1.614 &      0.065 &      0.408 &      0.014 &      0.223 &       0.02 &      0.146 &      0.011 &      0.035 &      0.002 &      0.019 &      0.002 &          -- &          -- &       1.25 \\
J042918+770911 &      4.702 &      0.188 &      0.945 &      0.032 &      0.517 &      0.044 &      0.266 &      0.015 &      0.085 &      0.004 &      0.026 &      0.003 &          -- &          -- &       1.47 \\
J044545+783856 &      1.807 &      0.072 &       0.65 &      0.022 &       0.46 &      0.041 &      0.292 &      0.022 &      0.156 &      0.004 &      0.113 &      0.013 &      0.068 &       0.01 &       0.78 \\
J050731+791257 &      1.859 &      0.074 &      0.629 &      0.021 &      0.496 &      0.044 &      0.307 &      0.019 &      0.155 &      0.004 &      0.122 &      0.014 &      0.066 &       0.01 &       0.77 \\
J050842+843204 &      0.149 &      0.007 &      0.295 &       0.01 &      0.302 &      0.031 &      0.242 &      0.019 &      0.246 &      0.008 &       0.25 &      0.027 &      0.169 &      0.025 &      -0.15 \\
J061837+782123 &      2.511 &      0.101 &      1.075 &      0.036 &      0.733 &      0.066 &      0.501 &      0.038 &      0.227 &      0.012 &      0.119 &      0.014 &          -- &          -- &       0.86 \\
J062205+871948 &      2.105 &      0.084 &      0.645 &      0.022 &      0.534 &      0.048 &      0.251 &      0.019 &        0.1 &      0.006 &      0.072 &      0.008 &          -- &          -- &       0.95 \\
J062602+820225 &      0.199 &      0.009 &      0.681 &      0.023 &      0.962 &      0.086 &      0.947 &      0.054 &      0.763 &      0.019 &      0.717 &      0.074 &      0.521 &       0.06 &      -0.36 \\
J063012+763245 &      2.187 &      0.088 &      0.783 &      0.026 &        0.7 &      0.063 &      0.339 &      0.025 &      0.152 &      0.009 &      0.092 &      0.011 &          -- &          -- &        0.9 \\
J063012+763245 &      1.512 &      0.061 &      0.481 &      0.016 &          -- &          -- &      0.193 &      0.015 &      0.085 &      0.004 &      0.053 &      0.006 &          -- &          -- &       0.95 \\
J063825+841106 &      3.834 &      0.153 &        1.1 &      0.037 &      0.737 &      0.066 &      0.426 &      0.029 &      0.197 &      0.005 &      0.155 &      0.018 &          -- &          -- &       0.91 \\
J064045+781327 &      1.605 &      0.064 &      0.688 &      0.023 &      0.403 &      0.036 &      0.249 &      0.018 &      0.092 &      0.004 &      0.048 &      0.005 &          -- &          -- &       0.99 \\
J064558+775502 &      2.054 &      0.082 &      0.608 &       0.02 &      0.458 &      0.041 &          -- &          -- &      0.098 &      0.006 &      0.067 &      0.008 &          -- &          -- &       0.97 \\
J071452+815153 &      1.704 &      0.068 &      0.551 &      0.018 &      0.419 &      0.037 &      0.255 &      0.016 &      0.141 &      0.004 &      0.127 &      0.014 &          -- &          -- &       0.73 \\
J072611+791130 &      0.092 &      0.005 &      0.501 &      0.017 &       0.75 &      0.057 &      0.904 &      0.051 &      0.818 &       0.02 &      0.742 &      0.077 &       0.48 &      0.054 &      -0.59 \\
J073433+765813 &      1.361 &      0.055 &      0.475 &      0.016 &      0.282 &      0.025 &      0.217 &      0.013 &      0.098 &      0.006 &      0.074 &      0.009 &          -- &          -- &       0.82 \\
J074XXX+802XXX &      9.574 &      0.383 &      3.322 &      0.112 &      2.408 &      0.185 &      1.699 &      0.124 &      0.954 &       0.05 &      0.681 &      0.076 &          -- &          -- &       0.75 \\
J075058+824158 &      3.741 &       0.15 &      1.845 &      0.062 &      1.579 &      0.119 &      1.421 &       0.08 &      0.959 &      0.024 &      0.881 &      0.091 &      0.643 &      0.095 &       0.41 \\
J080626+812620 &      1.506 &       0.06 &      0.406 &      0.014 &      0.251 &      0.023 &      0.127 &      0.008 &      0.045 &      0.002 &      0.033 &      0.004 &          -- &          -- &       1.08 \\
J080734+784610 &      1.574 &      0.063 &      0.525 &      0.018 &       0.39 &      0.035 &      0.218 &      0.013 &      0.106 &      0.003 &      0.084 &       0.01 &          -- &          -- &       0.83 \\
J082550+765313 &      1.903 &      0.076 &      0.689 &      0.023 &      0.491 &      0.044 &      0.303 &      0.018 &      0.159 &      0.004 &      0.121 &      0.014 &      0.028 &      0.004 &       0.78 \\
J083236+800601 &      2.019 &      0.081 &      0.867 &      0.029 &      0.638 &      0.057 &      0.392 &      0.023 &      0.184 &      0.005 &      0.132 &      0.015 &      0.051 &      0.008 &       0.77 \\
J084833+783003 &      5.369 &      0.215 &      1.509 &      0.051 &       0.86 &      0.077 &      0.508 &      0.029 &      0.203 &      0.006 &      0.127 &      0.014 &      0.038 &      0.006 &       1.06 \\
J085834+750121 &      0.239 &       0.01 &      0.948 &      0.032 &      0.578 &      0.052 &      0.302 &      0.017 &      0.097 &      0.006 &      0.048 &      0.005 &          -- &          -- &       0.45 \\
J090112+780930 &      1.422 &      0.057 &      0.447 &      0.015 &      0.252 &      0.023 &      0.165 &      0.012 &      0.069 &      0.003 &      0.059 &      0.007 &          -- &          -- &        0.9 \\
J090842+834543 &      1.039 &      0.042 &      0.448 &      0.015 &      0.325 &      0.029 &      0.221 &      0.013 &      0.118 &      0.003 &      0.099 &      0.011 &          -- &          -- &       0.66 \\
J092016+862845 &      2.122 &      0.085 &      0.519 &      0.017 &      0.345 &      0.031 &      0.173 &      0.013 &      0.065 &      0.003 &      0.036 &      0.004 &          -- &          -- &       1.15 \\
J093239+790629 &      9.234 &      0.369 &      2.241 &      0.075 &      1.333 &      0.119 &      0.667 &      0.038 &      0.235 &      0.006 &      0.127 &      0.014 &          -- &          -- &       1.21 \\
J093817+781528 &      2.062 &      0.083 &      0.699 &      0.023 &      0.426 &      0.038 &      0.272 &      0.016 &      0.125 &      0.004 &      0.075 &      0.009 &          -- &          -- &       0.94 \\
J093923+831526 &     11.592 &      0.464 &      2.953 &      0.099 &      1.802 &      0.134 &      0.917 &      0.052 &      0.304 &      0.009 &      0.171 &       0.02 &          -- &          -- &       1.19 \\
J094440+825408 &      1.904 &      0.076 &      0.735 &      0.025 &       0.54 &      0.048 &      0.358 &      0.023 &      0.173 &      0.006 &      0.129 &      0.015 &          -- &          -- &       0.76 \\
J095559+791134 &      1.295 &      0.052 &      0.415 &      0.014 &      0.254 &      0.023 &      0.152 &      0.009 &      0.061 &      0.003 &      0.047 &      0.005 &          -- &          -- &       0.94 \\
J100005+812702 &      2.541 &      0.102 &      0.864 &      0.029 &      0.561 &       0.05 &      0.304 &      0.017 &      0.122 &      0.007 &      0.069 &      0.008 &          -- &          -- &       1.02 \\
J100741+813150 &      3.632 &      0.145 &      0.855 &      0.029 &      0.496 &      0.044 &      0.257 &      0.015 &      0.096 &      0.003 &       0.05 &      0.006 &          -- &          -- &       1.21 \\
J100949+810719 &        1.5 &       0.06 &      0.455 &      0.015 &      0.328 &      0.029 &      0.195 &      0.012 &      0.081 &      0.004 &      0.062 &      0.007 &          -- &          -- &        0.9 \\
J101015+825014 &      0.741 &       0.03 &      0.504 &      0.017 &      0.507 &      0.045 &      0.622 &      0.035 &      0.607 &      0.015 &      0.628 &      0.066 &      0.548 &      0.064 &       0.05 \\
J101037+765052 &      3.076 &      0.123 &      0.806 &      0.027 &      0.522 &      0.047 &      0.235 &       0.02 &      0.071 &      0.004 &      0.033 &      0.004 &          -- &          -- &       1.28 \\
J101XXX+855XXX &      1.973 &      0.079 &       0.81 &      0.027 &      0.537 &      0.045 &      0.303 &      0.034 &      0.109 &      0.007 &      0.066 &      0.008 &          -- &          -- &       0.96 \\
J101734+810517 &      2.894 &      0.116 &      1.172 &      0.039 &      0.838 &      0.075 &      0.512 &       0.03 &      0.194 &      0.006 &       0.12 &      0.014 &          -- &          -- &        0.9 \\
J102326+803255 &      6.905 &      0.276 &      1.807 &      0.061 &      1.144 &      0.102 &      0.627 &      0.036 &      0.185 &      0.005 &      0.093 &      0.011 &          -- &          -- &       1.22 \\
J102926+785241 &      2.559 &      0.102 &      1.111 &      0.037 &      0.741 &      0.066 &      0.525 &       0.03 &       0.25 &      0.006 &      0.164 &      0.019 &          -- &          -- &       0.78 \\
J104423+805439 &      0.643 &      0.026 &      0.828 &      0.028 &      0.829 &      0.074 &      0.961 &      0.054 &      1.212 &      0.029 &      1.362 &       0.14 &      1.206 &      0.134 &      -0.21 \\
J105150+791341 &      1.894 &      0.076 &      0.525 &      0.018 &      0.314 &      0.028 &      0.179 &       0.01 &      0.078 &      0.004 &       0.06 &      0.007 &          -- &          -- &       0.98 \\
J110405+793253 &      0.338 &      0.014 &      0.514 &      0.017 &      0.491 &      0.044 &      0.377 &      0.021 &      0.237 &      0.006 &      0.172 &      0.018 &       0.11 &      0.016 &       0.19 \\
J110412+765859 &      7.554 &      0.302 &      2.341 &      0.079 &      1.518 &      0.113 &      0.887 &       0.05 &      0.395 &       0.01 &      0.224 &      0.023 &          -- &          -- &       0.99 \\
J111342+765449 &      1.534 &      0.061 &      0.471 &      0.016 &       0.29 &      0.026 &      0.183 &      0.011 &      0.095 &      0.005 &      0.066 &      0.007 &          -- &          -- &       0.89 \\
J112342+773123 &      1.516 &      0.061 &      0.407 &      0.014 &      0.228 &       0.02 &      0.151 &      0.009 &      0.055 &      0.003 &      0.036 &      0.004 &          -- &          -- &       1.06 \\
J114829+782721 &       0.97 &      0.039 &       0.42 &      0.014 &        0.3 &      0.027 &      0.206 &      0.012 &      0.088 &      0.005 &      0.074 &      0.009 &          -- &          -- &       0.73 \\
J115312+805829 &       1.39 &      0.056 &      1.343 &      0.045 &      1.887 &      0.146 &      1.922 &      0.108 &      1.715 &      0.042 &      1.696 &      0.175 &      1.251 &      0.138 &      -0.06 \\
J115504+753439 &      1.822 &      0.073 &      0.817 &      0.027 &      0.594 &      0.053 &      0.407 &      0.025 &      0.201 &      0.006 &      0.137 &      0.016 &          -- &          -- &       0.73 \\
J115522+815709 &      2.017 &      0.081 &      0.671 &      0.023 &      0.428 &      0.038 &      0.227 &      0.019 &        0.1 &      0.004 &      0.065 &      0.007 &          -- &          -- &       0.97 \\
J115608+823505 &        1.4 &      0.056 &      0.404 &      0.014 &       0.25 &      0.023 &      0.175 &       0.02 &      0.065 &      0.003 &      0.056 &      0.007 &          -- &          -- &       0.91 \\
J115713+811824 &      1.269 &      0.051 &      0.981 &      0.033 &      0.766 &       0.06 &      0.556 &      0.033 &      0.301 &      0.009 &      0.222 &      0.023 &          -- &          -- &       0.49 \\
J122015+792732 &      1.409 &      0.056 &      0.517 &      0.017 &      0.334 &       0.03 &      0.214 &      0.012 &      0.084 &      0.004 &      0.043 &      0.005 &          -- &          -- &       0.99 \\
J122340+804004 &      1.204 &      0.048 &      0.705 &      0.024 &      0.701 &      0.063 &      0.752 &      0.043 &      0.767 &      0.019 &      0.808 &      0.083 &      0.775 &      0.087 &       0.11 \\
J122518+860839 &      1.532 &      0.061 &      0.453 &      0.015 &      0.294 &      0.026 &      0.164 &      0.016 &      0.082 &      0.004 &      0.041 &      0.005 &          -- &          -- &       1.02 \\
J123708+835704 &      1.909 &      0.076 &      0.778 &      0.026 &      0.667 &      0.051 &      0.333 &       0.02 &      0.186 &      0.005 &      0.122 &      0.014 &          -- &          -- &       0.78 \\
J125736+834231 &      1.145 &      0.046 &      0.475 &      0.016 &      0.375 &      0.034 &       0.27 &      0.019 &      0.176 &      0.005 &      0.164 &      0.019 &          -- &          -- &       0.55 \\
J130035+805438 &      4.837 &      0.194 &      1.251 &      0.042 &      0.777 &      0.069 &      0.386 &      0.022 &      0.166 &      0.004 &      0.107 &      0.012 &          -- &          -- &       1.08 \\
J130538+815626 &      1.734 &      0.069 &       0.49 &      0.016 &       0.32 &      0.029 &       0.17 &       0.01 &      0.077 &      0.004 &      0.054 &      0.006 &          -- &          -- &       0.98 \\
J130609+800825 &      2.218 &      0.089 &      0.785 &      0.026 &      0.597 &      0.053 &      0.365 &      0.022 &       0.18 &      0.007 &      0.142 &      0.016 &          -- &          -- &       0.78 \\
J130705+764918 &      1.178 &      0.047 &      0.754 &      0.025 &      0.531 &      0.048 &      0.359 &      0.021 &       0.18 &      0.006 &      0.124 &      0.014 &          -- &          -- &       0.64 \\
J130811+854424 &      1.711 &      0.069 &      0.593 &       0.02 &      0.418 &      0.037 &      0.236 &      0.015 &      0.156 &      0.009 &      0.108 &      0.012 &          -- &          -- &       0.78 \\
J131723+821916 &      3.126 &      0.125 &      0.869 &      0.029 &       0.46 &      0.041 &       0.29 &      0.017 &      0.114 &      0.007 &       0.08 &      0.009 &          -- &          -- &       1.04 \\
J132053+845011 &      0.702 &      0.028 &      0.436 &      0.015 &      0.484 &      0.044 &       0.28 &      0.021 &      0.211 &      0.007 &      0.189 &      0.022 &          -- &          -- &       0.37 \\
J132145+831613 &      0.829 &      0.033 &      0.565 &      0.019 &      0.485 &      0.043 &      0.333 &      0.025 &      0.275 &      0.015 &      0.262 &       0.03 &          -- &          -- &       0.33 \\
J132331+780947 &      1.192 &      0.048 &      0.482 &      0.016 &      0.346 &      0.031 &       0.23 &      0.014 &      0.106 &      0.003 &      0.072 &      0.008 &          -- &          -- &       0.79 \\
J132351+794251 &      0.574 &      0.023 &      0.599 &       0.02 &      0.642 &      0.057 &      0.575 &      0.033 &      0.476 &      0.012 &      0.435 &      0.046 &      0.331 &      0.042 &       0.08 \\
J135639+794340 &      2.121 &      0.085 &      0.579 &      0.019 &      0.349 &      0.031 &      0.189 &      0.011 &      0.072 &      0.004 &      0.036 &      0.004 &          -- &          -- &       1.15 \\
J135755+764320 &      0.419 &      0.017 &      0.647 &      0.022 &      0.575 &      0.051 &        0.7 &       0.04 &      0.769 &       0.02 &      0.819 &      0.085 &      0.761 &      0.085 &      -0.19 \\
J141419+790547 &      1.445 &      0.058 &      0.424 &      0.014 &      0.251 &      0.023 &      0.153 &       0.01 &      0.082 &      0.004 &      0.056 &      0.007 &          -- &          -- &       0.92 \\
J141718+805939 &      1.519 &      0.061 &      0.541 &      0.018 &      0.356 &      0.032 &      0.217 &      0.016 &      0.115 &      0.005 &      0.077 &      0.009 &          -- &          -- &       0.84 \\
J141947+760033 &      3.216 &      0.129 &      0.981 &      0.033 &      0.635 &      0.057 &      0.388 &      0.022 &      0.192 &      0.007 &      0.137 &      0.016 &      0.058 &      0.009 &       0.89 \\
J142248+770416 &      1.094 &      0.044 &      0.541 &      0.018 &      0.329 &      0.029 &      0.173 &       0.01 &      0.063 &      0.003 &      0.029 &      0.003 &          -- &          -- &       1.03 \\
J142613+794607 &      1.257 &       0.05 &      0.407 &      0.014 &      0.266 &      0.024 &      0.168 &      0.012 &      0.067 &      0.003 &      0.036 &      0.004 &          -- &          -- &          1 \\
J143547+760526 &      2.644 &      0.106 &      1.304 &      0.044 &      0.979 &      0.087 &      0.724 &      0.041 &      0.428 &      0.011 &      0.323 &      0.034 &      0.191 &      0.028 &       0.59 \\
J144314+770726 &      6.648 &      0.266 &      1.882 &      0.063 &      1.069 &      0.095 &      0.552 &      0.034 &      0.225 &      0.006 &      0.131 &      0.015 &          -- &          -- &       1.11 \\
J144709+765621 &      6.168 &      0.247 &      1.667 &      0.056 &      0.997 &      0.089 &      0.559 &      0.032 &      0.226 &      0.006 &      0.144 &      0.016 &          -- &          -- &       1.06 \\
J150008+751851 &      3.858 &      0.154 &      0.784 &      0.026 &      0.395 &      0.035 &       0.19 &      0.011 &      0.076 &      0.004 &      0.044 &      0.005 &          -- &          -- &       1.26 \\
J150207+860811 &      0.863 &      0.035 &      0.416 &      0.014 &      0.379 &      0.034 &      0.229 &      0.014 &      0.108 &      0.003 &      0.087 &       0.01 &          -- &          -- &       0.65 \\
J151304+814326 &      2.144 &      0.086 &      0.782 &      0.026 &      0.554 &       0.05 &      0.302 &      0.022 &      0.134 &      0.004 &      0.088 &       0.01 &          -- &          -- &        0.9 \\
J153112+770604 &      1.446 &      0.058 &      0.566 &      0.019 &      0.346 &      0.031 &       0.18 &       0.01 &      0.056 &      0.003 &      0.021 &      0.002 &          -- &          -- &        1.2 \\
J153700+815431 &        0.2 &      0.009 &      0.433 &      0.015 &       0.39 &      0.035 &      0.314 &      0.018 &        0.2 &      0.005 &      0.163 &      0.019 &      0.095 &      0.014 &       0.06 \\
J160222+801558 &      4.507 &       0.18 &      1.016 &      0.034 &      0.515 &      0.046 &      0.255 &      0.015 &          -- &          -- &      0.049 &      0.005 &          -- &          -- &       1.28 \\
J160929+793954 &      2.806 &      0.112 &      1.239 &      0.042 &      0.965 &      0.086 &      0.711 &       0.04 &      0.431 &      0.012 &      0.333 &      0.035 &      0.193 &      0.028 &        0.6 \\
J161940+854921 &      3.965 &      0.159 &      1.643 &      0.055 &      1.322 &      0.098 &      0.801 &      0.046 &      0.433 &      0.011 &      0.319 &      0.034 &          -- &          -- &       0.71 \\
J163051+823345 &          -- &          -- &      0.875 &      0.029 &      0.378 &      0.034 &      0.259 &      0.019 &      0.086 &      0.005 &      0.052 &      0.006 &          -- &          -- &       1.36 \\
J163226+823220 &          -- &          -- &      0.802 &      0.027 &      0.777 &      0.064 &      0.644 &      0.037 &      0.555 &      0.014 &      0.615 &      0.065 &      0.647 &      0.096 &       0.13 \\
J163925+863153 &      2.234 &      0.089 &      0.852 &      0.029 &      0.633 &      0.057 &      0.418 &      0.026 &      0.233 &      0.006 &      0.173 &      0.019 &          -- &          -- &       0.72 \\
J164843+754628 &      5.799 &      0.232 &      1.942 &      0.065 &      1.246 &      0.111 &      0.746 &      0.042 &      0.365 &       0.01 &      0.245 &      0.026 &      0.109 &      0.016 &       0.89 \\
J165752+792808 &      2.458 &      0.098 &      0.872 &      0.029 &      0.599 &      0.054 &      0.399 &      0.024 &      0.179 &      0.006 &      0.123 &      0.014 &          -- &          -- &       0.85 \\
J171416+761245 &      1.916 &      0.077 &      0.459 &      0.015 &      0.268 &      0.024 &      0.139 &      0.008 &      0.056 &      0.003 &      0.031 &      0.004 &          -- &          -- &       1.17 \\
J172359+765312 &      0.216 &      0.009 &      0.424 &      0.014 &      0.342 &       0.03 &      0.358 &       0.02 &      0.343 &      0.011 &      0.315 &      0.033 &      0.279 &      0.042 &      -0.11 \\
J172529+770805 &      0.979 &      0.039 &      0.565 &      0.019 &      0.425 &      0.038 &      0.262 &      0.016 &      0.134 &      0.005 &      0.096 &      0.011 &          -- &          -- &       0.66 \\
J172550+772624 &      3.077 &      0.123 &      1.163 &      0.039 &      0.795 &      0.071 &      0.499 &      0.028 &       0.27 &      0.007 &      0.206 &      0.023 &      0.132 &      0.019 &       0.76 \\
J173021+794916 &      4.444 &      0.178 &       1.02 &      0.034 &      0.648 &      0.058 &      0.309 &      0.033 &      0.126 &      0.004 &      0.064 &      0.007 &          -- &          -- &        1.2 \\
J173734+844543 &      1.971 &      0.079 &      0.444 &      0.015 &      0.258 &      0.023 &      0.128 &      0.009 &      0.054 &      0.003 &      0.029 &      0.003 &          -- &          -- &       1.19 \\
J175056+814736 &      1.139 &      0.046 &       0.44 &      0.015 &      0.269 &      0.024 &      0.161 &      0.011 &      0.054 &      0.003 &      0.033 &      0.004 &          -- &          -- &          1 \\
J180045+782804 &      1.918 &      0.077 &      2.224 &      0.075 &      2.679 &      0.205 &      2.883 &      0.162 &      2.674 &      0.067 &       2.69 &      0.278 &      2.427 &      0.272 &       -0.1 \\
J183712+851449 &      1.852 &      0.074 &       0.69 &      0.023 &      0.526 &      0.047 &      0.246 &      0.016 &      0.113 &      0.004 &      0.057 &      0.007 &          -- &          -- &       0.98 \\
J184XXX+794XXX &     30.187 &      1.207 &     12.944 &      0.435 &       9.44 &      0.859 &       5.94 &      0.416 &       3.63 &       0.37 &          -- &          -- &          -- &          -- &       0.85 \\
J184502+765230 &      1.517 &      0.061 &      0.536 &      0.018 &      0.347 &      0.031 &      0.251 &      0.016 &      0.101 &      0.006 &      0.061 &      0.007 &          -- &          -- &       0.91 \\
J185750+774636 &      1.636 &      0.066 &      0.474 &      0.016 &      0.291 &      0.026 &      0.166 &       0.01 &      0.074 &      0.004 &      0.042 &      0.005 &          -- &          -- &       1.03 \\
J190350+853647 &      2.533 &      0.101 &      0.905 &       0.03 &       0.63 &      0.056 &      0.411 &       0.03 &      0.192 &      0.005 &      0.147 &      0.017 &          -- &          -- &        0.8 \\
J190919+781330 &      0.823 &      0.033 &      0.465 &      0.016 &       0.33 &       0.03 &      0.202 &      0.012 &      0.094 &      0.003 &      0.057 &      0.007 &          -- &          -- &       0.75 \\
J193419+795606 &      2.721 &      0.109 &      0.765 &      0.026 &      0.518 &      0.046 &      0.231 &      0.016 &      0.101 &      0.006 &      0.049 &      0.005 &          -- &          -- &       1.13 \\
J193739+835629 &      0.844 &      0.034 &       0.43 &      0.014 &      0.392 &      0.035 &      0.239 &      0.015 &       0.22 &      0.008 &      0.265 &      0.028 &          -- &          -- &       0.33 \\
J194136+850138 &      2.477 &      0.099 &      0.662 &      0.022 &      0.596 &      0.053 &       0.29 &       0.02 &      0.129 &      0.007 &      0.094 &      0.011 &          -- &          -- &       0.92 \\
J194340+785829 &      1.107 &      0.044 &      0.499 &      0.017 &      0.444 &       0.04 &      0.205 &      0.014 &      0.111 &      0.004 &      0.078 &      0.009 &          -- &          -- &       0.75 \\
J194420+781602 &      1.631 &      0.065 &      0.503 &      0.017 &      0.322 &      0.029 &      0.147 &      0.009 &      0.044 &      0.002 &          -- &          -- &          -- &          -- &       1.45 \\
J194958+765413 &      1.827 &      0.073 &       0.51 &      0.017 &      0.312 &      0.028 &      0.164 &      0.009 &      0.074 &      0.004 &      0.051 &      0.006 &          -- &          -- &       1.01 \\
J200531+775243 &      0.806 &      0.032 &      1.061 &      0.036 &      1.343 &      0.102 &      1.463 &      0.083 &      1.453 &      0.035 &      1.428 &      0.147 &      1.324 &      0.146 &      -0.16 \\
J202235+761126 &      0.567 &      0.023 &      0.429 &      0.014 &        0.5 &      0.045 &      0.449 &      0.025 &      0.414 &       0.01 &      0.393 &      0.041 &      0.369 &      0.055 &        0.1 \\
J2042XX+750XXX &          -- &          -- &      1.144 &      0.038 &      1.172 &        0.1 &      0.842 &      0.062 &      0.555 &      0.028 &          -- &          -- &          -- &          -- &       0.71 \\
J204541+762510 &      3.238 &       0.13 &      0.953 &      0.032 &      0.603 &      0.054 &      0.474 &      0.027 &      0.257 &      0.008 &      0.201 &      0.023 &          -- &          -- &       0.79 \\
J205033+752622 &      2.107 &      0.084 &      0.556 &      0.019 &      0.365 &      0.032 &      0.182 &      0.012 &      0.082 &      0.004 &      0.046 &      0.005 &          -- &          -- &       1.08 \\
J210407+763307 &      15.52 &      0.621 &      3.891 &      0.131 &       2.27 &      0.169 &      1.314 &      0.074 &      0.486 &      0.012 &      0.259 &      0.027 &          -- &          -- &       1.16 \\
J2118XX+751XXX &      4.717 &      0.189 &      1.261 &      0.042 &       0.66 &      0.059 &      0.419 &      0.027 &      0.135 &      0.004 &      0.078 &      0.009 &          -- &          -- &       1.16 \\
J211956+765734 &      1.188 &      0.048 &      0.432 &      0.015 &      0.307 &      0.027 &       0.15 &      0.011 &      0.082 &      0.004 &       0.05 &      0.006 &          -- &          -- &       0.89 \\
J212926+845326 &      3.514 &      0.141 &      1.274 &      0.043 &      0.809 &      0.072 &      0.482 &      0.035 &      0.168 &      0.007 &      0.072 &      0.008 &          -- &          -- &        1.1 \\
J213008+835730 &      5.098 &      0.204 &      1.798 &       0.06 &      1.332 &      0.119 &      0.846 &      0.048 &      0.382 &       0.01 &      0.279 &      0.032 &          -- &          -- &       0.82 \\
J213139+843011 &      0.445 &      0.018 &      0.677 &      0.023 &      0.647 &      0.058 &      0.422 &      0.032 &      0.241 &      0.007 &      0.185 &      0.021 &          -- &          -- &       0.25 \\
J213334+823905 &       1.93 &      0.077 &      0.915 &      0.031 &      0.809 &      0.072 &      0.617 &      0.035 &      0.433 &       0.01 &       0.39 &      0.045 &      0.192 &      0.028 &       0.45 \\
J213929+833953 &      1.418 &      0.057 &       0.48 &      0.016 &      0.306 &      0.027 &      0.217 &      0.012 &      0.107 &      0.006 &      0.081 &      0.009 &          -- &          -- &       0.81 \\
J214928+754045 &      1.735 &      0.069 &      0.524 &      0.018 &      0.329 &      0.029 &      0.202 &      0.012 &      0.104 &      0.006 &      0.073 &      0.009 &          -- &          -- &        0.9 \\
J215657+833714 &      0.474 &      0.019 &      0.474 &      0.016 &      0.442 &      0.039 &      0.285 &      0.022 &      0.214 &      0.009 &      0.184 &      0.021 &          -- &          -- &       0.27 \\
J215712+764642 &      2.283 &      0.091 &      0.777 &      0.026 &       0.56 &       0.05 &      0.309 &      0.018 &      0.149 &      0.006 &      0.091 &      0.011 &          -- &          -- &       0.91 \\
J220955+835356 &      1.787 &      0.072 &      0.578 &      0.019 &        0.4 &      0.036 &      0.191 &      0.018 &       0.09 &      0.005 &      0.069 &      0.008 &          -- &          -- &       0.92 \\
J222800+753219 &      2.039 &      0.082 &      0.651 &      0.022 &      0.429 &      0.038 &      0.289 &      0.018 &      0.142 &      0.005 &      0.111 &      0.013 &          -- &          -- &       0.82 \\
J224714+855542 &      1.432 &      0.057 &      0.516 &      0.017 &      0.356 &      0.032 &      0.205 &      0.013 &      0.098 &      0.006 &      0.076 &      0.009 &          -- &          -- &       0.83 \\
J230122+795406 &      1.447 &      0.058 &      0.431 &      0.014 &      0.291 &      0.026 &      0.154 &       0.01 &      0.077 &      0.004 &      0.044 &      0.005 &          -- &          -- &       0.99 \\
J230138+820015 &      1.518 &      0.061 &       0.45 &      0.015 &      0.378 &      0.034 &      0.196 &      0.016 &      0.085 &      0.004 &      0.046 &      0.005 &          -- &          -- &       0.99 \\
J232503+791715 &      0.705 &      0.028 &      1.136 &      0.038 &      0.912 &      0.082 &      0.578 &      0.037 &      0.285 &      0.007 &      0.189 &      0.021 &          -- &          -- &       0.37 \\
J232640+823158 &      2.964 &      0.119 &      1.001 &      0.034 &      0.728 &      0.065 &       0.35 &      0.022 &       0.16 &      0.009 &      0.066 &      0.007 &          -- &          -- &       1.07 \\
J232803+761738 &      1.466 &      0.059 &      0.459 &      0.015 &       0.26 &      0.023 &      0.172 &       0.01 &      0.072 &      0.004 &      0.053 &      0.006 &          -- &          -- &       0.94 \\
J234403+822640 &      5.667 &      0.227 &      3.777 &      0.127 &      2.904 &      0.259 &      1.787 &      0.102 &      0.896 &      0.022 &       0.65 &      0.067 &      0.292 &      0.043 &       0.61 \\
J234914+751744 &      1.428 &      0.057 &       0.43 &      0.014 &      0.322 &      0.029 &        0.2 &      0.013 &      0.117 &      0.007 &      0.096 &      0.011 &          -- &          -- &       0.76 \\
J235413+804753 &      1.634 &      0.065 &      0.482 &      0.016 &      0.307 &      0.027 &      0.178 &       0.01 &      0.075 &      0.004 &      0.058 &      0.007 &          -- &          -- &       0.94 \\
J2355XX+795XXX &      6.463 &      0.259 &      1.706 &      0.057 &      0.967 &      0.073 &      0.603 &      0.034 &      0.196 &      0.007 &      0.094 &      0.011 &          -- &          -- &        1.2 \\
J235622+815252 &      0.569 &      0.023 &      0.521 &      0.017 &      0.454 &      0.041 &        0.5 &      0.032 &      0.586 &      0.016 &      0.667 &      0.071 &      0.713 &       0.08 &      -0.04 \\
\end{supertabular}  

\end{onecolumn}}}
\end{center}

\begin{center}
\begin{onecolumn}
   \begin{figure*}
   \resizebox{16cm}{!}{\includegraphics{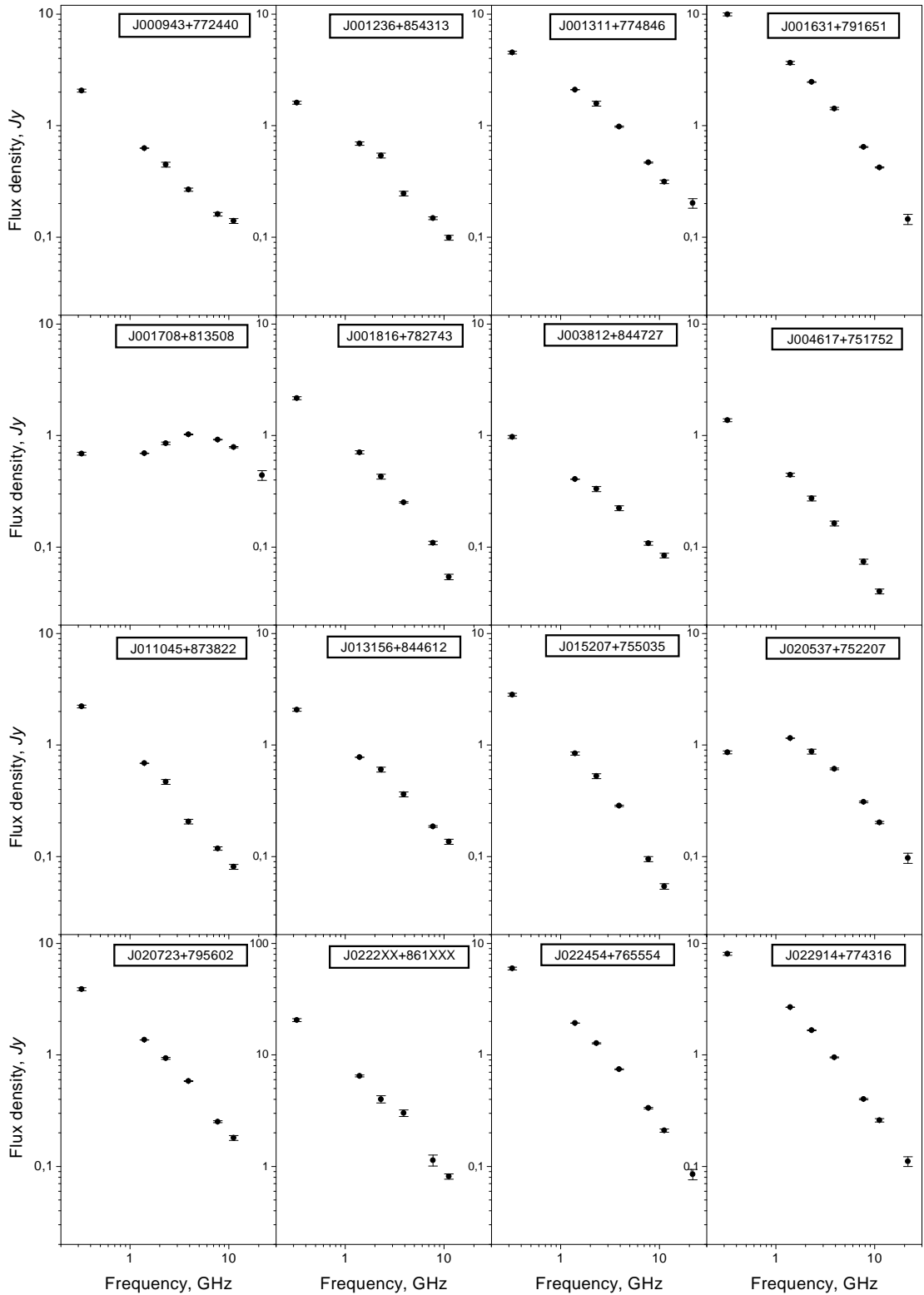}}
   \hfill
   \caption{Spectra of the sources observed during RATAN-600 NCP survey. 
            WENSS and NVSS data included for comparison
            at 0.325 GHz and 1.4 GHz.}
   \label{Spectra}
   \end{figure*}

\setcounter{figure}{6}

   \begin{figure*}
   \resizebox{16cm}{!}{\includegraphics{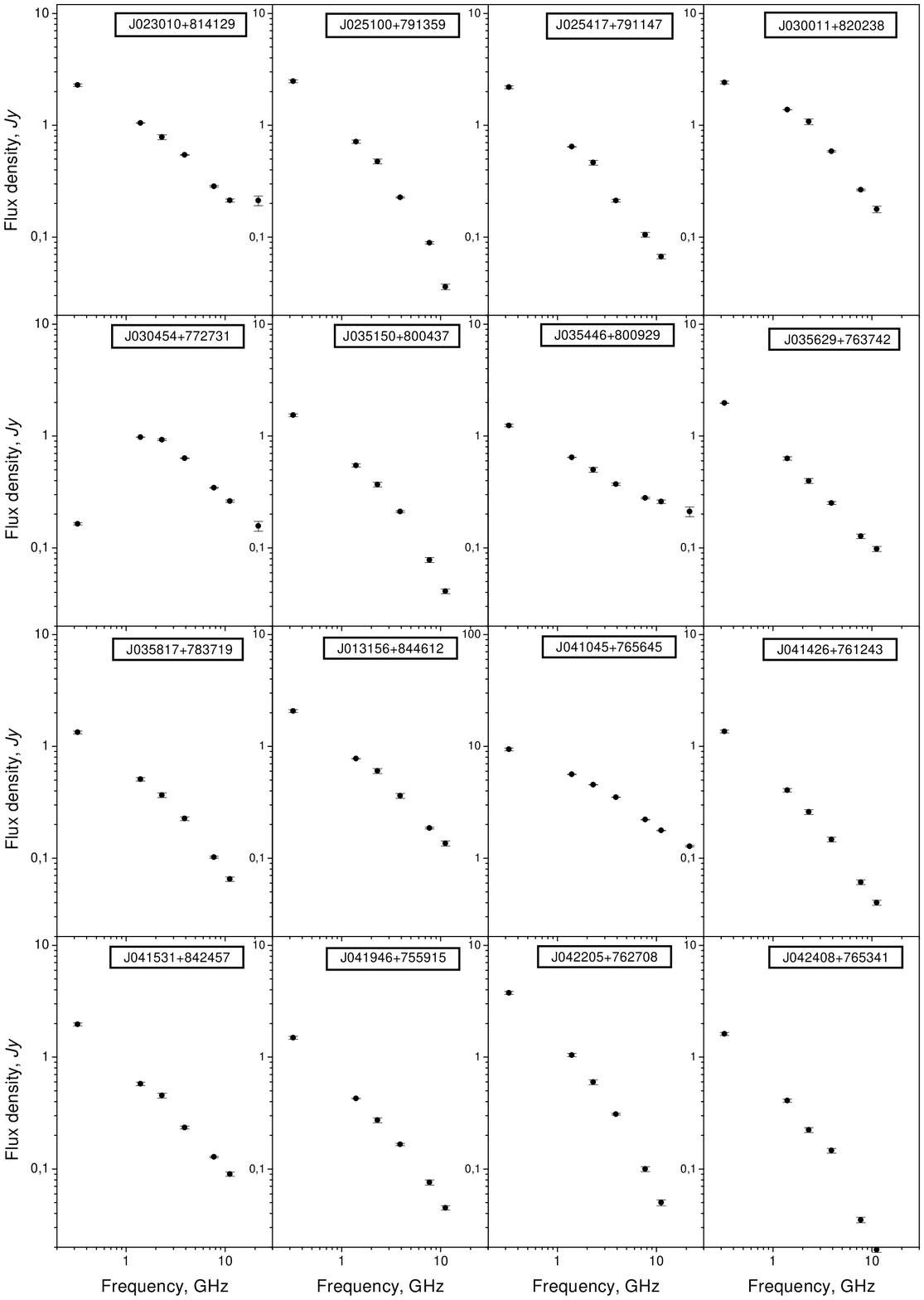}}
   \hfill
   \caption{Spectra of the sources observed during RATAN-600 NCP survey
(continued) }
   \end{figure*}

\setcounter{figure}{6}

   \begin{figure*}
   \resizebox{16cm}{!}{\includegraphics{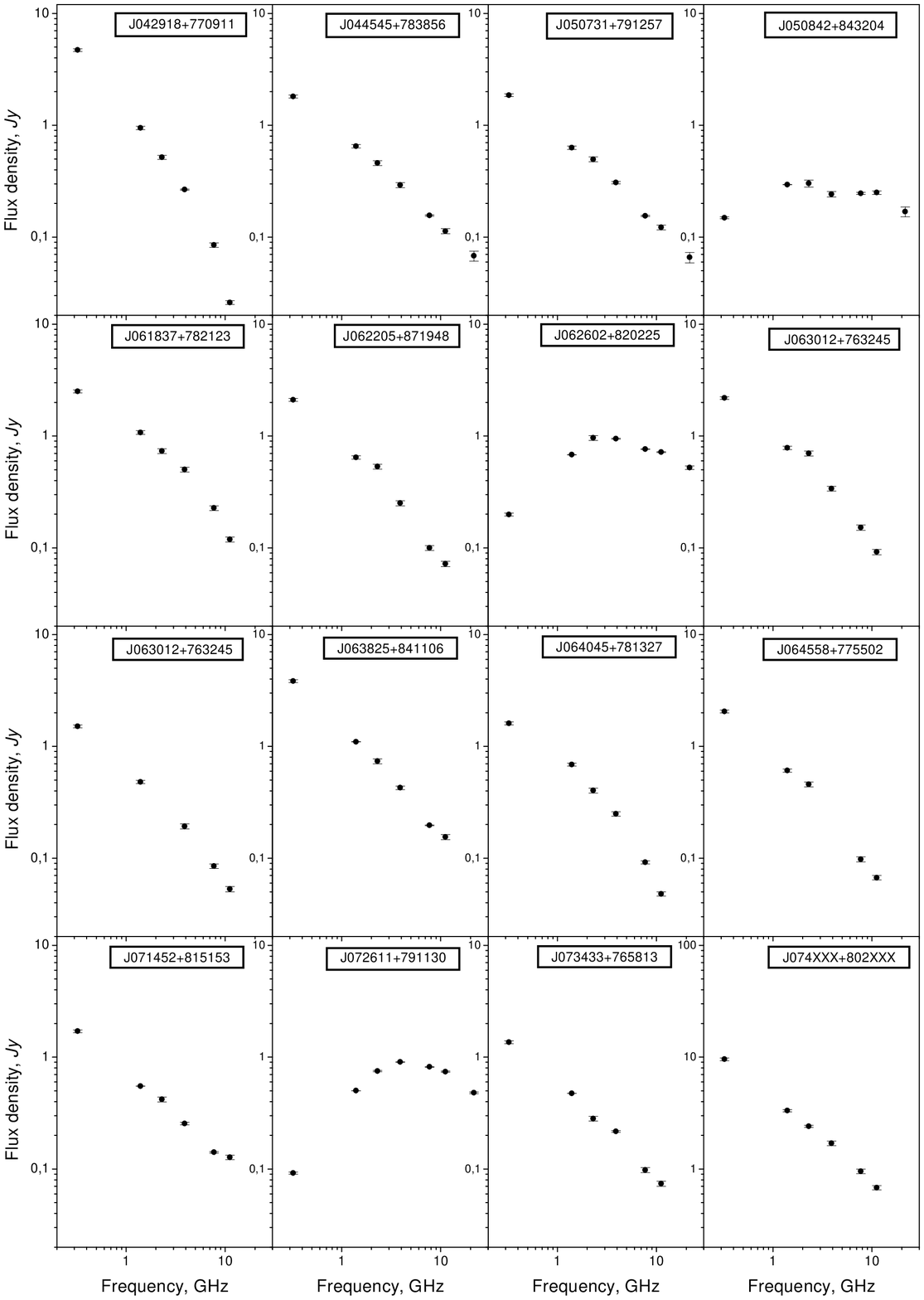}}
   \hfill
   \caption{Spectra of the sources observed during RATAN-600 NCP survey
(continued) }
   \end{figure*}

\setcounter{figure}{6}

   \begin{figure*}
   \resizebox{16cm}{!}{\includegraphics{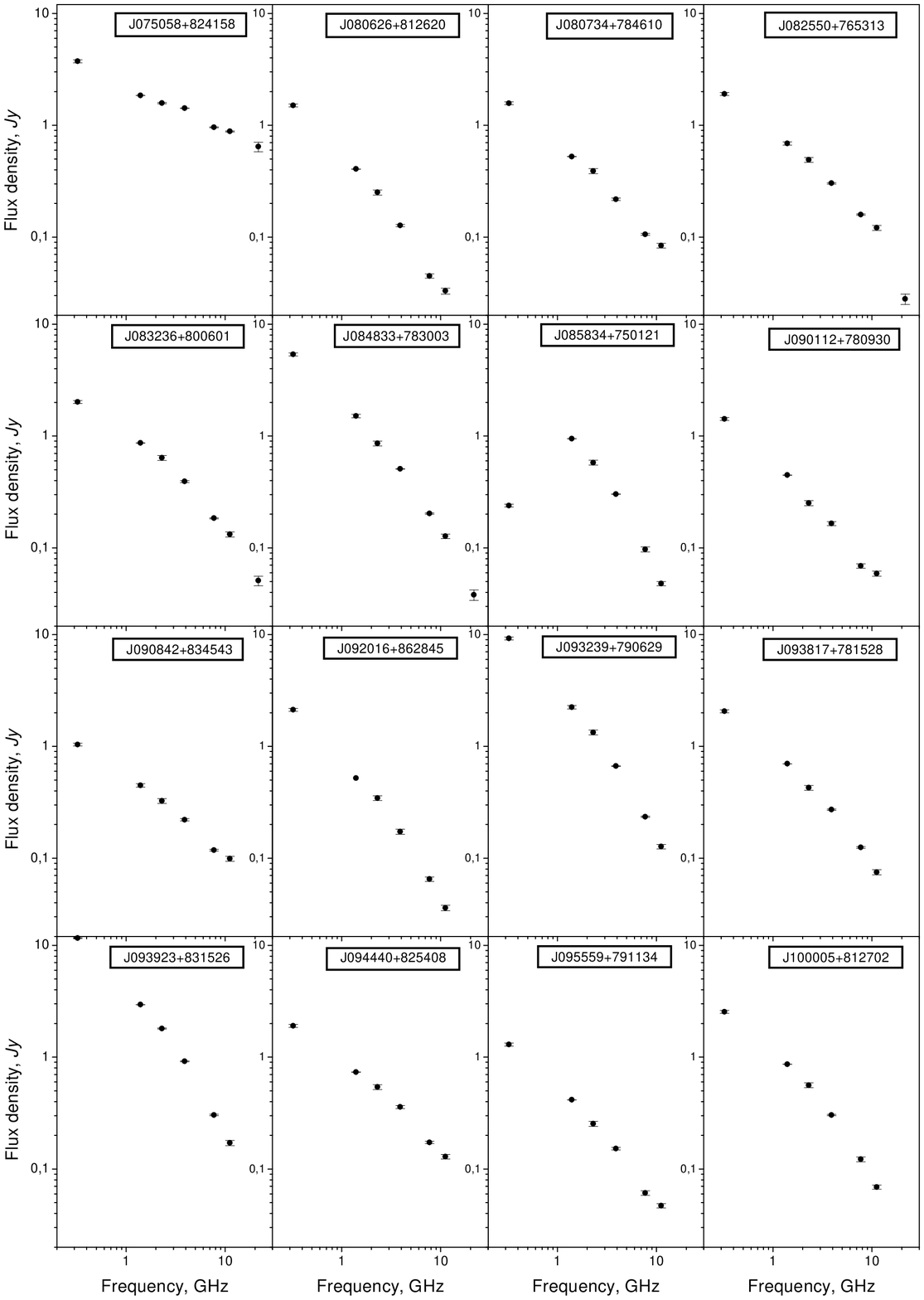}}
   \hfill
   \caption{Spectra of the sources observed during RATAN-600 NCP survey
(continued) }
   \end{figure*}

\setcounter{figure}{6}

   \begin{figure*}
   \resizebox{16cm}{!}{\includegraphics{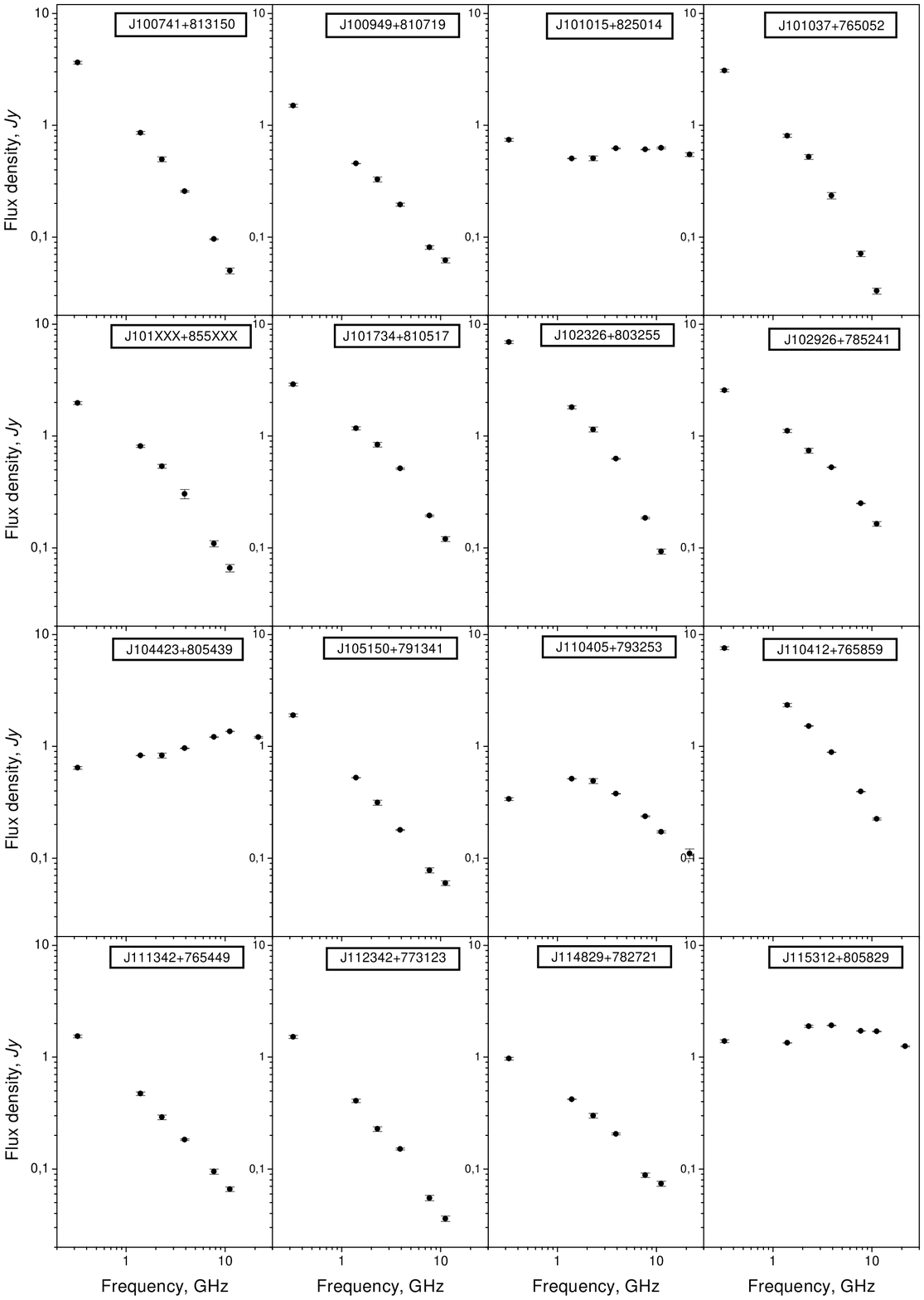}}
   \hfill
   \caption{Spectra of the sources observed during RATAN-600 NCP survey
(continued) }
   \end{figure*}

\setcounter{figure}{6}

   \begin{figure*}
   \resizebox{16cm}{!}{\includegraphics{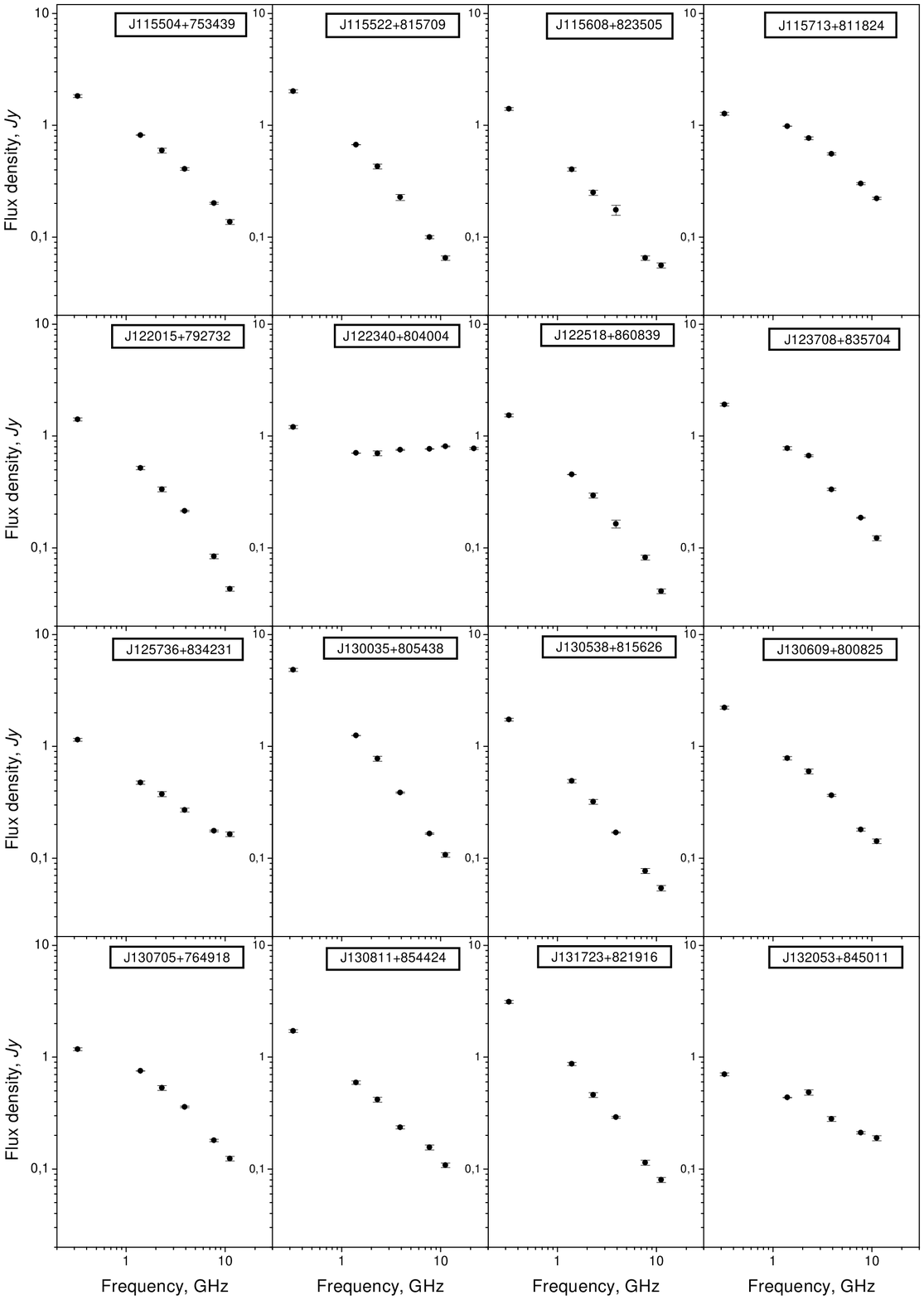}}
   \hfill
   \caption{Spectra of the sources observed during RATAN-600 NCP survey
(continued) }
   \end{figure*}

\setcounter{figure}{6}

   \begin{figure*}
   \resizebox{16cm}{!}{\includegraphics{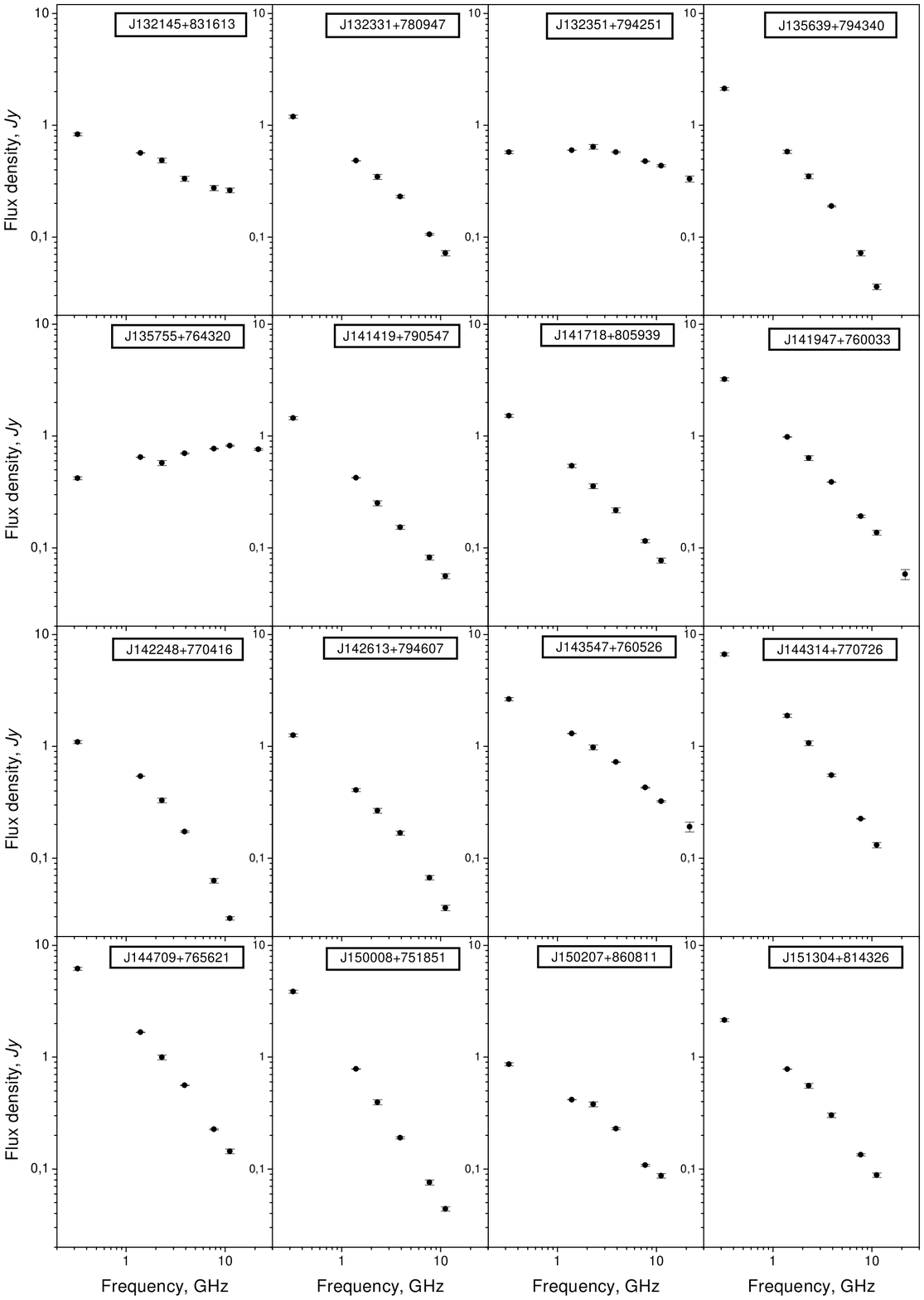}}
   \hfill
   \caption{Spectra of the sources observed during RATAN-600 NCP survey
(continued) }
   \end{figure*}

\setcounter{figure}{6}

   \begin{figure*}
   \resizebox{16cm}{!}{\includegraphics{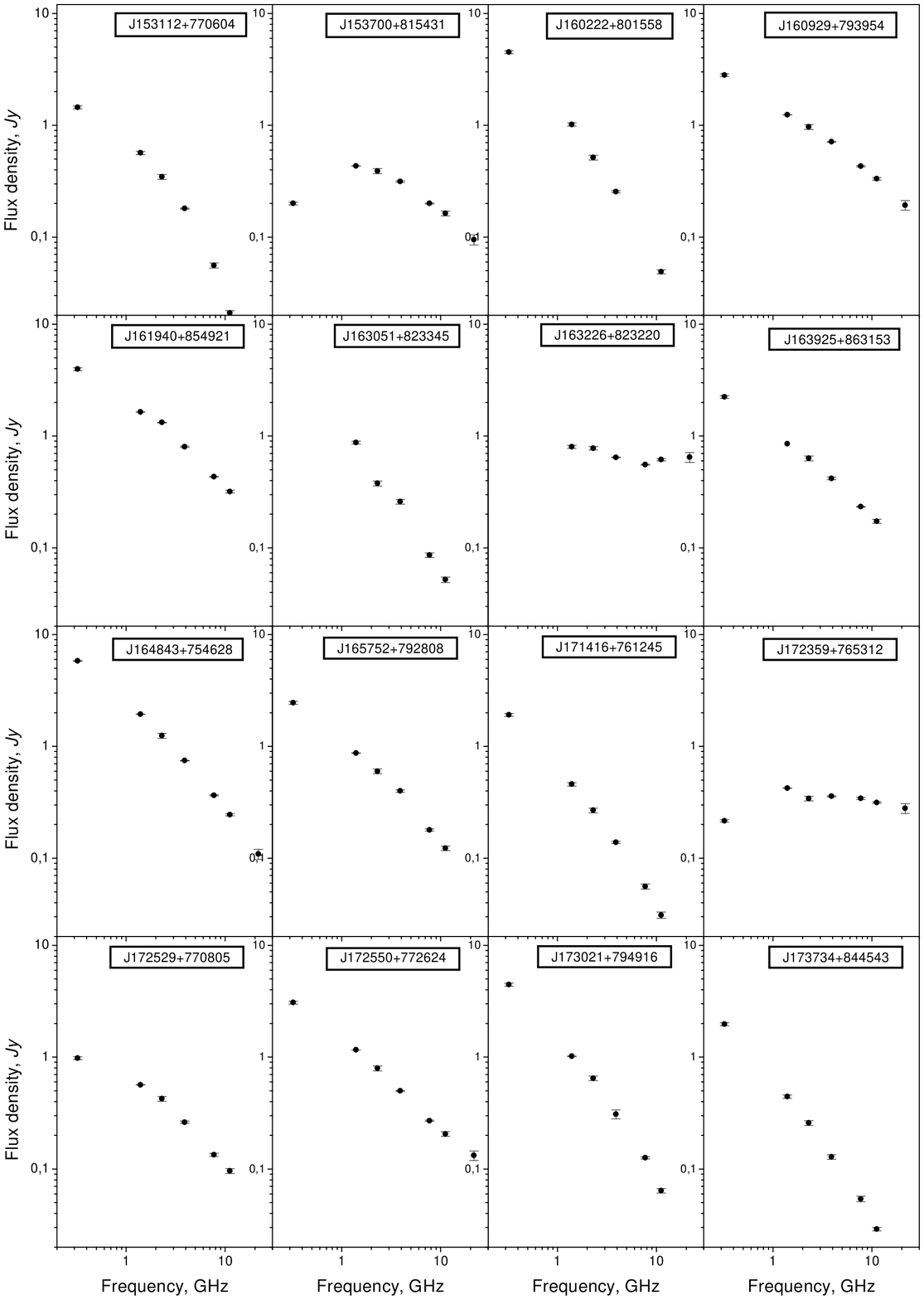}}
   \hfill
   \caption{Spectra of the sources observed during RATAN-600 NCP survey
(continued) }
   \end{figure*}

\setcounter{figure}{6}

   \begin{figure*}
   \resizebox{16cm}{!}{\includegraphics{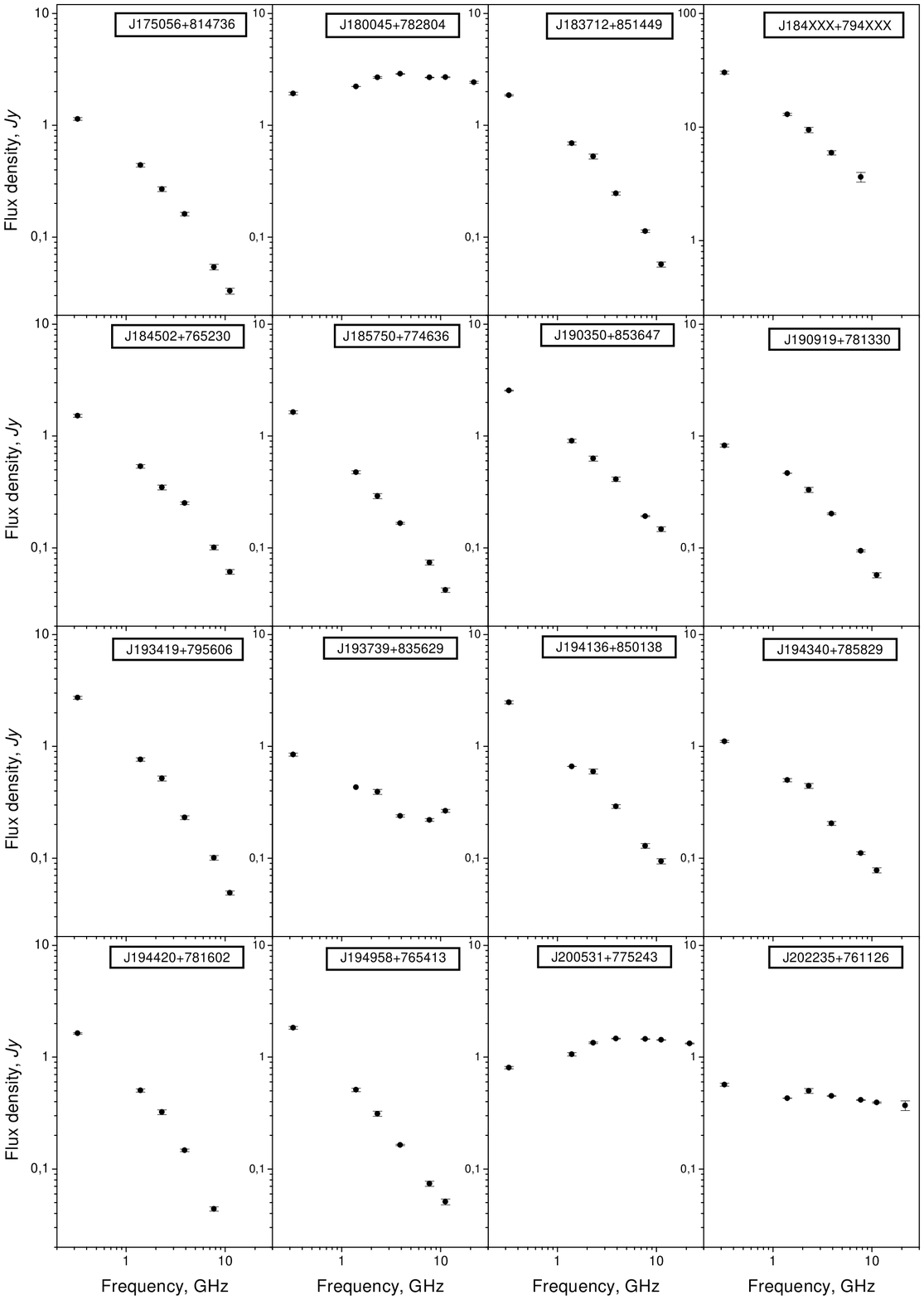}}
   \hfill
   \caption{Spectra of the sources observed during RATAN-600 NCP survey
(continued) }
   \end{figure*}

\setcounter{figure}{6}

   \begin{figure*}
   \resizebox{16cm}{!}{\includegraphics{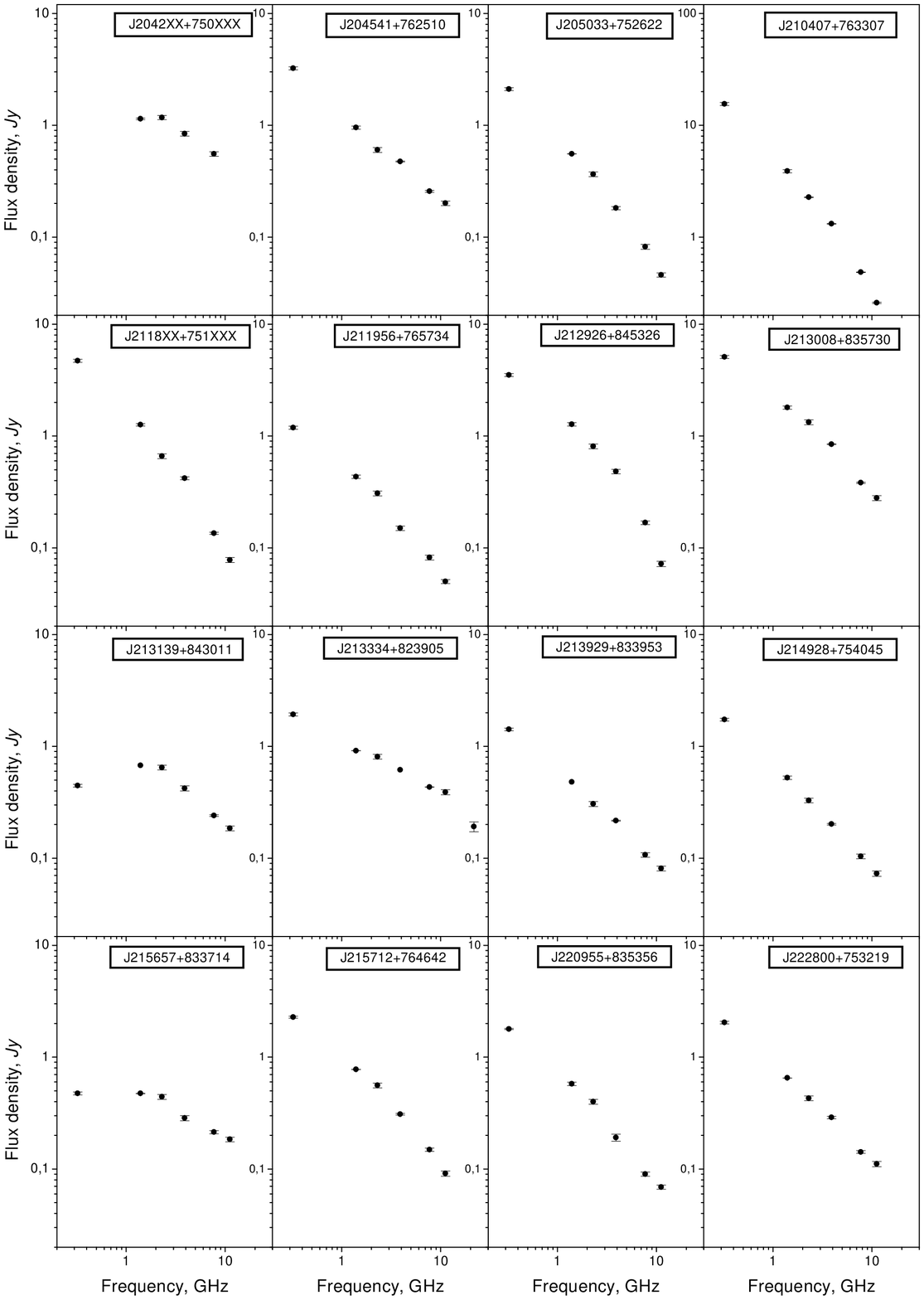}}
   \hfill
   \caption{Spectra of the sources observed during RATAN-600 NCP survey
(continued) }
   \end{figure*}

\setcounter{figure}{6}

   \begin{figure*}
   \resizebox{16cm}{!}{\includegraphics{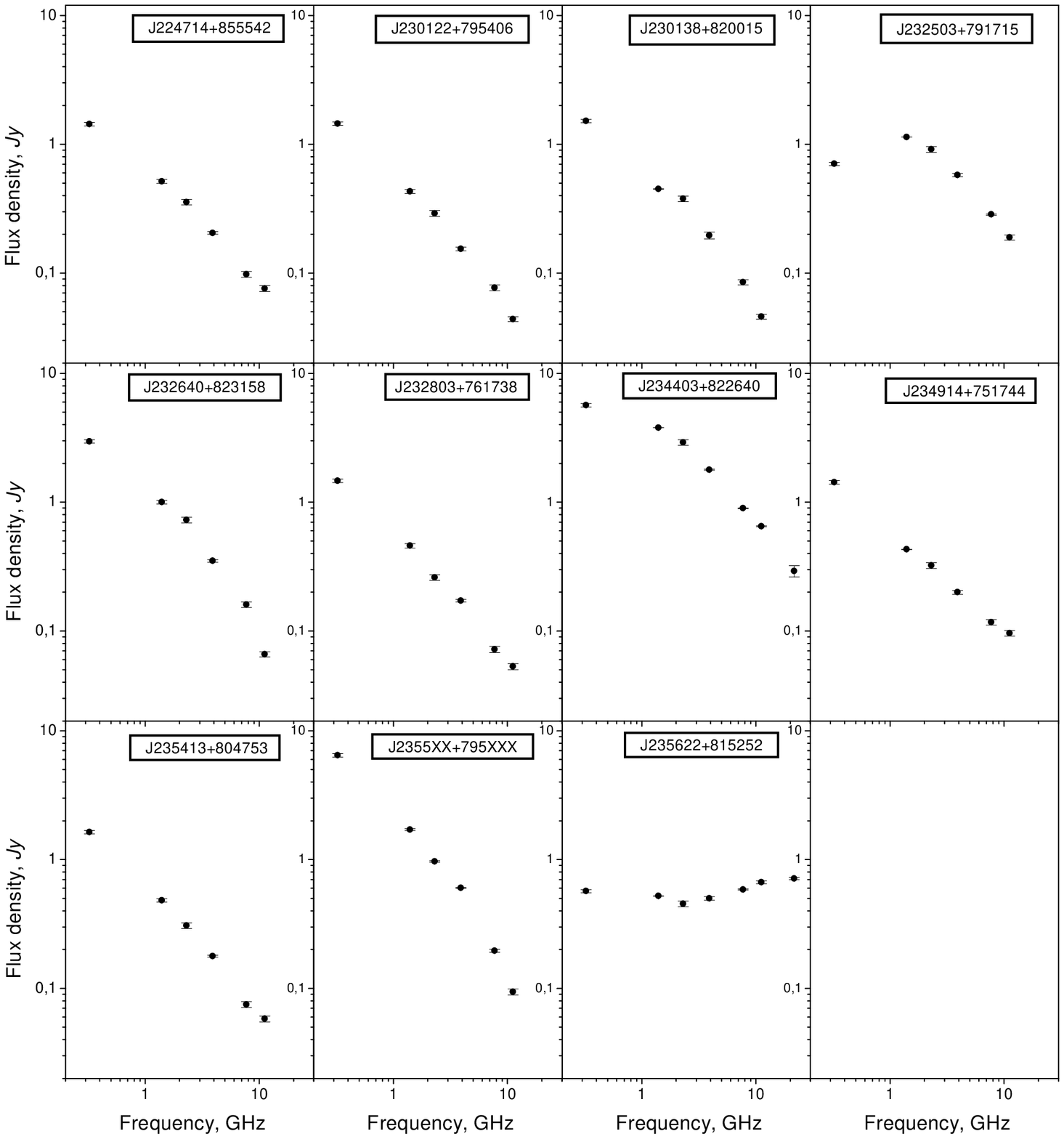}}
   \hfill
   \caption{Spectra of the sources observed during RATAN-600 NCP survey
(continued) }
   \end{figure*}

\end{onecolumn}
\end{center}

\end{document}